\documentclass[conference]{IEEEtran}
\IEEEoverridecommandlockouts

\usepackage{cite}
\usepackage{amsmath,amssymb,amsfonts}
\usepackage{graphicx}
\usepackage{textcomp}
\usepackage{multirow}
\usepackage{xcolor}
\usepackage{cleveref}
\usepackage{subcaption}
\usepackage{float}
\usepackage{pifont}
\usepackage{algorithm}
\usepackage{algpseudocode}
\usepackage{mathtools}
\usepackage{soul}
\usepackage{placeins}
\usepackage{enumitem}
\usepackage{xspace}
\usepackage{booktabs}
\usepackage{array}

\usepackage[hyphens]{url}

\def\BibTeX{{\rm B\kern-.05em{\sc i\kern-.025em b}\kern-.08em
    T\kern-.1667em\lower.7ex\hbox{E}\kern-.125emX}}
\begin{document}

\pdfpagewidth=8.5in
\pdfpageheight=11in

\captionsetup[subfigure]{labelformat=simple}
\renewcommand\thesubfigure{(\alph{subfigure})}
\crefname{figure}{figure}{figures}

\newcommand{\cmark}{\textcolor{green!60!black}{\ding{51}}}
\newcommand{\xmark}{\textcolor{red}{\ding{55}}}
\newcommand{\gmark}{\textcolor{gray}{\raisebox{0.5ex}{\rule{1.2ex}{0.25ex}}}}
\definecolor{jcred}{HTML}{e31a1c}
\definecolor{jcgreen}{HTML}{33a02c}
\definecolor{jcblue}{HTML}{1f78b4}
\definecolor{jcorange}{HTML}{ff7f00}
\definecolor{jcpurple}{HTML}{6a3d9a}
\definecolor{jcbrown}{HTML}{b15928}
\newcommand{\best}[1]{\textcolor{jcgreen}{\bf #1}}

\newcommand{\llama}{\textsc{Llama-}\scalebox{0.95}{2}\xspace}
\newcommand{\llamaIII}{\textsc{Llama-}\scalebox{0.95}{3}\xspace}
\newcommand{\llamaIIISmall}{\textsc{Llama-}\scalebox{0.9}{3-8B}\xspace}
\newcommand{\llamaIIIBig}{\textsc{Llama-}\scalebox{0.9}{3-70B}\xspace}
\newcommand{\llamaIVMVK}{\textsc{Llama-}\scalebox{0.9}{4-Maverick}\xspace}
\newcommand{\llamaSmall}{\textsc{Llama-}\scalebox{0.9}{2-7B}\xspace}
\newcommand{\llamaMedium}{\textsc{Llama-}\scalebox{0.9}{2-13B}\xspace}
\newcommand{\llamaIIIOneSMALL}{\textsc{Llama-}\scalebox{0.9}{3.1-8B}\xspace}
\newcommand{\llamaIIIPLUSBIG}{\textsc{Llama-}\scalebox{0.9}{3.3-70B}\xspace}
\newcommand{\llamaBig}{\textsc{Llama-}\scalebox{0.9}{2-70B}\xspace}
\newcommand{\llamaTiny}{\textsc{Llama\scalebox{0.9}{3.2-1B}} }
\newcommand{\llamafamily}{\textsc{Llama}\xspace}
\newcommand{\wiki}{WikiText-2\xspace}
\definecolor{darkblue}{RGB}{0, 70, 140}

\pagenumbering{arabic}

\title{Combating the Memory Walls: Optimization Pathways for Long-Context Agentic LLM Inference}
\author{
\IEEEauthorblockN{
Haoran Wu\textsuperscript{1},
Can Xiao\textsuperscript{2},
Jiayi Nie\textsuperscript{1},
Xuan Guo\textsuperscript{2},
Binglei Lou\textsuperscript{2},
Jeffrey T.H. Wong\textsuperscript{2},
Zhiwen Mo\textsuperscript{2},\\
Cheng Zhang\textsuperscript{2},
Przemyslaw Forys\textsuperscript{2},
Chengyang Ai\textsuperscript{3},
Timi Adeniran\textsuperscript{1},
Wayne Luk\textsuperscript{2},
Hongxiang Fan\textsuperscript{2},\\
Jianyi Cheng\textsuperscript{3},
Timothy M. Jones\textsuperscript{1},
Rika Antonova\textsuperscript{1},
Robert Mullins\textsuperscript{1},
Aaron Zhao\textsuperscript{2}
}
\IEEEauthorblockA{
\textsuperscript{1}University of Cambridge \quad
\textsuperscript{2}Imperial College London \quad
\textsuperscript{3}University of Edinburgh
}
}

\maketitle
\thispagestyle{plain}
\pagestyle{plain}


\begin{abstract}

LLMs now form the backbone of AI agents for a diverse array of applications, including tool use, command-line interfaces, and web or computer interaction. These agentic LLM inference tasks are fundamentally different from chatbot-focused inference
--- they often have much larger context lengths to capture complex, prolonged inputs, such as an entire webpage DOM or complicated tool call trajectories. 
This, in turn, generates significant off-chip memory traffic for hardware at the inference stage and causes the workload to be constrained by the two memory walls, namely the \textit{bandwidth} and \textit{capacity} walls, preventing the compute units from achieving high utilization.

In this paper, we introduce PLENA, a hardware–software co-designed system that applies three core optimization pathways. PLENA features a novel flattened systolic-array architecture (\textit{Pathway 1}) and efficient compute and memory units that support an asymmetric quantization scheme (\textit{Pathway 2}). It also provides native support for FlashAttention (\textit{Pathway 3}). In addition, PLENA is developed with a complete software–hardware stack, including a custom ISA, a compiler, a transaction-level simulator, and an automated design-space exploration flow. Experimental results show that PLENA delivers up to 2.23$\times$ and 4.70$\times$ higher throughput than the A100 GPU and TPU v6e, respectively, under identical multiplier counts and memory configurations during LLaMA agentic inference. PLENA also achieves up to 4.04$\times$ higher energy efficiency than A100 GPU. The full PLENA system—including its simulator, compiler, ISA, and RTL implementation—will be open-sourced to the research community.

\end{abstract}

\section{Introduction}

Transformers have revolutionized AI across various fields, including language, vision, and science~\cite{wei2022emergentabilitieslargelanguage,attention,kojima2023largelanguagemodelszeroshot}. Transformer-based autoregressive large language models (LLMs), like GPT~\cite{gpt4} and LLaMA~\cite{llama}, are now widely deployed in many applications, such as chatbots~\cite{chatgpt}, code generation~\cite{code_gen}, tool-use and computer-use workflows \cite{agentic_workflow}. 

\begin{figure}[t]
  \centering

  \begin{subfigure}{\linewidth}
    \centering
    \includegraphics[width=\linewidth,height=0.30\textheight,keepaspectratio]{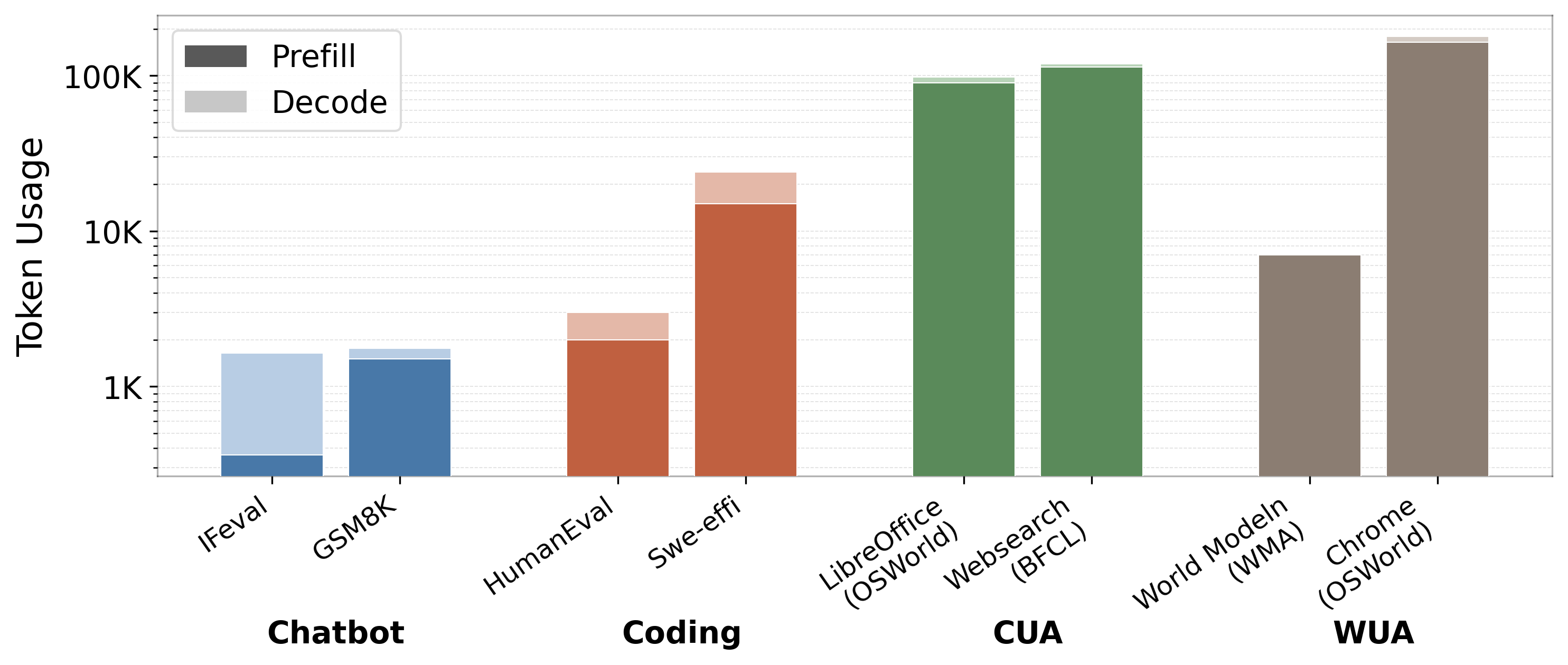}
    \caption{Token usage comparison across standard chatbot~\cite{ifeval,gsm8k}, coding~\cite{humaneval, sweeffi}, and agentic workloads, including Computer Use Agent (CUA)~\cite{osworld, agent_s2} and Web Use Agent (WUA)~\cite{webagent, osworld}.}
    \label{fig:Workload_analysis}
  \end{subfigure}

  \medskip

  \begin{subfigure}{0.48\linewidth}
    \centering
    \includegraphics[width=\linewidth,height=0.22\textheight,keepaspectratio]{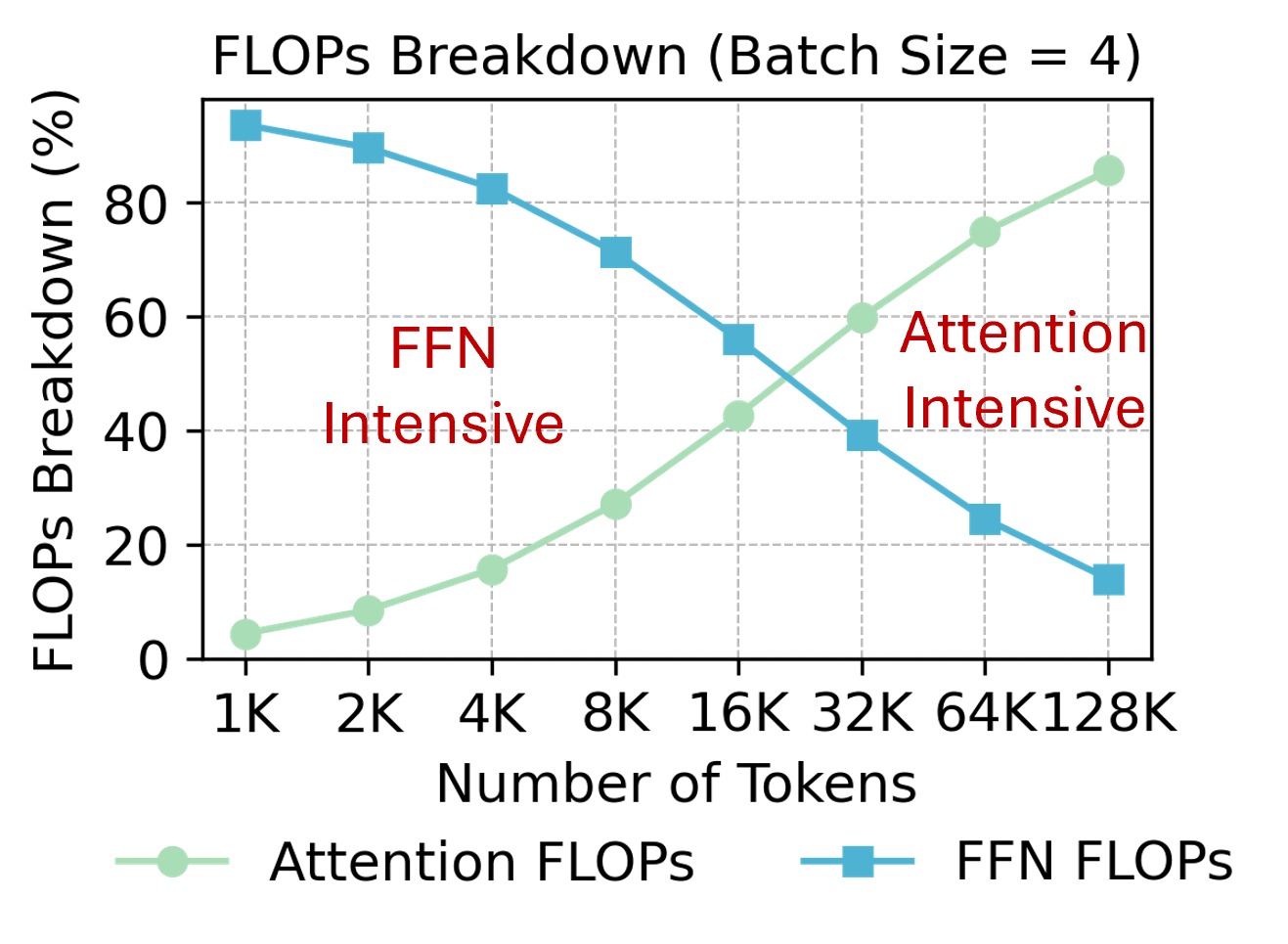}
    \caption{Compute intensity shifts from FFN to Attention blocks with an increasing context length.}
    \label{fig:Flops_breakdown}
  \end{subfigure}\hfill
  \begin{subfigure}{0.48\linewidth}
    \centering
    \includegraphics[width=\linewidth,height=0.22\textheight,keepaspectratio]{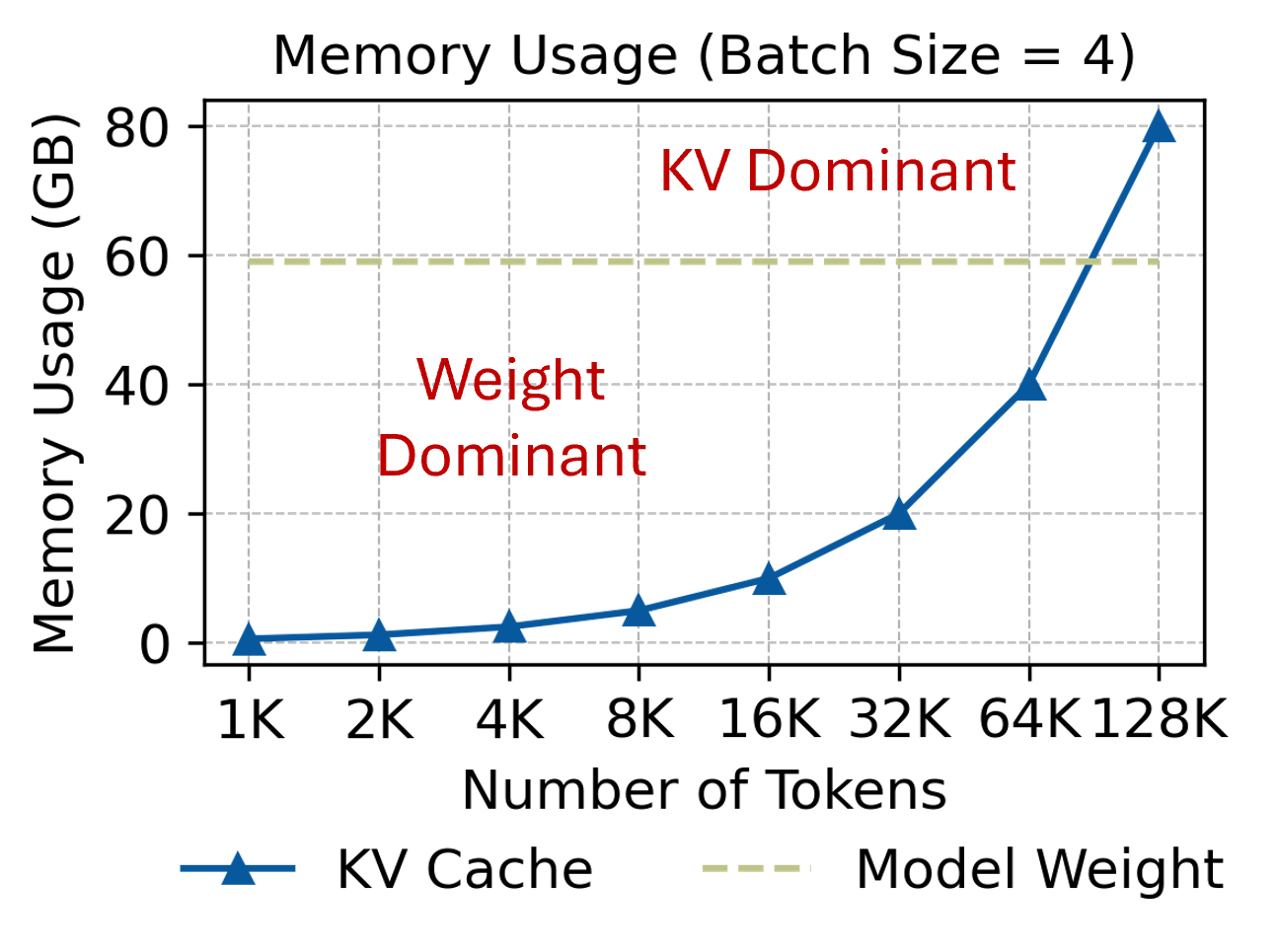}
    \caption{KV cache scales with context length, eventually dominates memory usage.}
    \label{fig:mem_utilization}
  \end{subfigure}
  \caption{An illustration of agentic inference workloads shows how they typically generate many more tokens per single inference run \subref{fig:Workload_analysis}, contain both FFN-compute-intensive and attention-compute-intensive phases \subref{fig:Flops_breakdown}, and include weight memory-capacity-dominant and KV-dominant phases \subref{fig:mem_utilization} within a single inference run.}
  \label{fig:two-layer-subfig}
  \vspace{-10pt}
\end{figure}

The rapid rise of agentic capabilities of  LLMs, e.g.~computer-use~\cite{OmniParser}, tool-use~\cite{browser_use2024, WebVoyager}, and command-line agents~\cite{command_agent}, heavily relies on their ability to process and reason over very large context windows. For instance, command-line agents need to both comprehend and generate large-scale codebases~\cite{code_gen_long_context, multi-swe, LongCodeBench}, while tool- and computer-use agentic workflows must keep track of multiple pieces of information across prolonged inputs---such as a complete web page DOM---which typically require very long contexts \cite{webdom_long_context,workarena,BrowserGym}.
\Cref{fig:Workload_analysis} shows that, when compared to chatbot workloads, agentic workloads consume~100$\times$ more tokens per inference on average and up to~1,000$\times$ in extreme. In response, modern LLMs have delibrately expanded their context windows: the original GPT-3~\cite{gpt3} supports roughly 2K tokens, whereas GPT-4\cite{gpt4} reaches up to 32K tokens, and \llamaIVMVK\cite{llama4_maverick} to 1M tokens.

To clarify the computational impact of agentic workloads, \Cref{fig:Flops_breakdown} analyzes a \llamaIIIBig model with long-context capability and demonstrates that, when the number of generated tokens is small, the Feed-Forward Networks (FFNs) contribute most of the inference FLOPs. As the number of generated tokens increases, however, the attention layers gradually dominate FLOP counts. Because inference is autoregressive, these two computational phases naturally coexist within a single long-context decoding run. For example, in the LongWriter~\cite{longwriter} workload, the prefilling phase completes at around 5K tokens, after which the decoding phase expands the context to a large value -- 85K tokens. As the sequence grows, the computational intensity (in FLOPs) start to transition from FFN-dominated to attention-dominated, with the crossover occurring at roughly 19K generated tokens, as shown in \Cref{fig:Flops_breakdown}. This shift makes both FFN and attention layers practical and necessary targets for optimization.

Agentic LLM inference also consumes significant HBM resources.
\Cref{fig:mem_utilization} identifies two major limiting factors on the memory side. 
First, the large number of KV values and weights that must be read, together with the portion of KV values written back, impose substantial memory bandwidth demands. 
Second, as context length increases, the KV-cache requirement grows linearly, quickly increasing memory usage and often surpassing the size of the model weights, making HBM capacity a primary limiting factor.
For example, in \llamaIIIBig, at a 128k context~\cite{llama3}, the FP16 KV cache for a single batch is approximately 39 GB, which limits how many batches can be kept on the chip~\cite{AI_and_Mem_Wall}.
Building on this observation, we suggest that there are two main challenges on the off-chip memory side, namely, (i) the limited memory bandwidth and (ii) the restricted memory capacity.
We collectively term these \emph{memory walls}. Together, they prevent devices from reaching peak performance at inference time, consistent with observations in prior work~\cite{AI_and_Mem_Wall, Efficient_LLM_Inference, LLMCompass}.

The memory wall phenomenon leads to the underutilization of computing resources on modern hardware, such as TPUs and GPUs. This effect is particularly evident in compute units dedicated to General Matrix-Matrix Multiplication (GEMM) operations ($\mathbb{R}^{M \times K} \times \mathbb{R}^{K \times N} \to \mathbb{R}^{M \times N}$), denoted as $(M, K) \times (K, N)$, which constitute the core computational workload during LLM inference~\cite{Transistive}.
At the microarchitectural level, most hardware is built with square-shaped systolic arrays or matrix multiplication units, typically designed so that the $M$ and $N$ dimensions are close in size to $K$. For example, TPU v3~\cite{TPU_V3_Systolic_Array} features a 128×128 systolic array, supporting $M = K = N = 128$ GEMM operations. 
However, in long-context agentic models, as shown in \Cref{fig:mem_utilization}, memory often becomes the primary constrain for the inference batch size. This results in a \textit{fat GEMM} operation, where the batch-related dimension (typically $M$ in $(M,K)\times(K,N)$) is much smaller than the other operating dimension. This essentially produces an uneven matrix shape\footnote{All KVs must be stored, so the batch size (the \(M\) dimension) is kept lower than the hidden size (\(K\)). While various offloading techniques are available~\cite{deepspeedinference}, they complicate system-level trade-offs and tend to make the system more memory I/O--bound.}. This imbalance hinders systolic arrays and Tensor Cores from achieving a high computational resources utilization rate~\cite{FlashDecoding}.

To this end, we propose the \textbf{P}rogrammable \textbf{L}ong-context \textbf{E}fficient \textbf{N}eural \textbf{A}ccelerator (PLENA), an efficient transformer model accelerator system designed to maintain high utilization of GEMM units across all inference stages (prefilling and decoding), particularly for agentic LLM inference tasks with large contexts. PLENA achieves high efficiency for long-context inference by exploring three optimization pathways across both hardware and software design spaces.

\begin{figure}[t]
  \centering
  \begin{subfigure}[t]{0.32\linewidth}
    \centering
    \includegraphics[width=\linewidth]{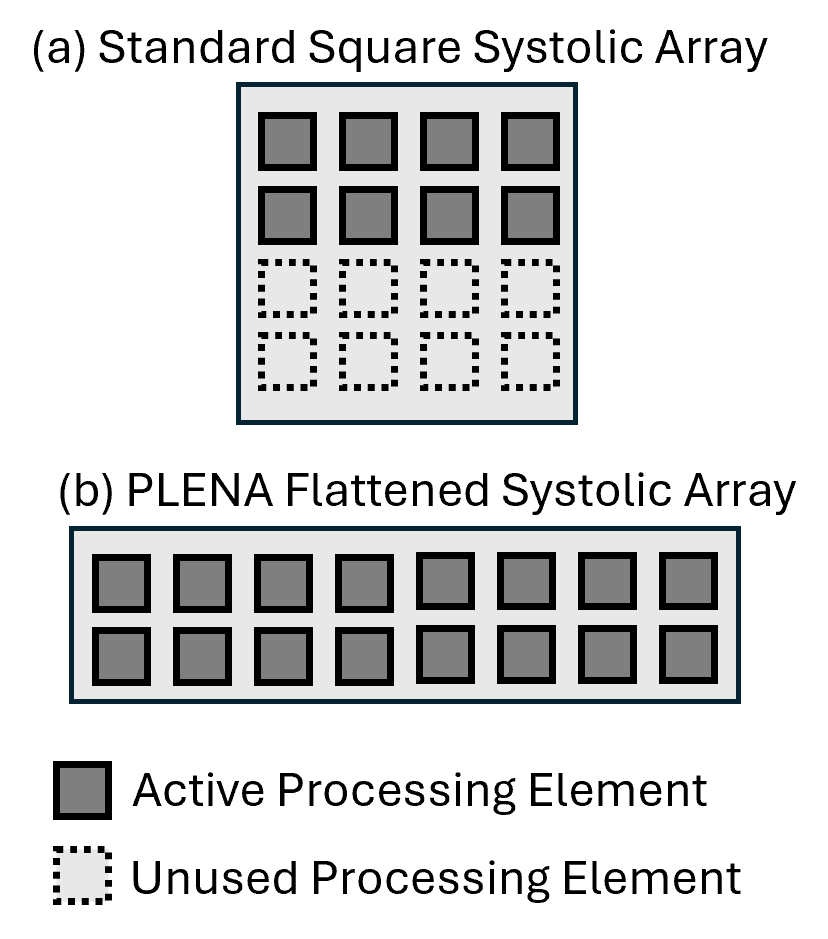}
    \caption{PLENA achieves higher utilization than the standard square systolic array with the same resources.}
    \label{fig:sa_compare_plot}
  \end{subfigure}
  \hfill
  \begin{subfigure}[t]{0.65\linewidth}
    \centering
    \includegraphics[width=\linewidth]{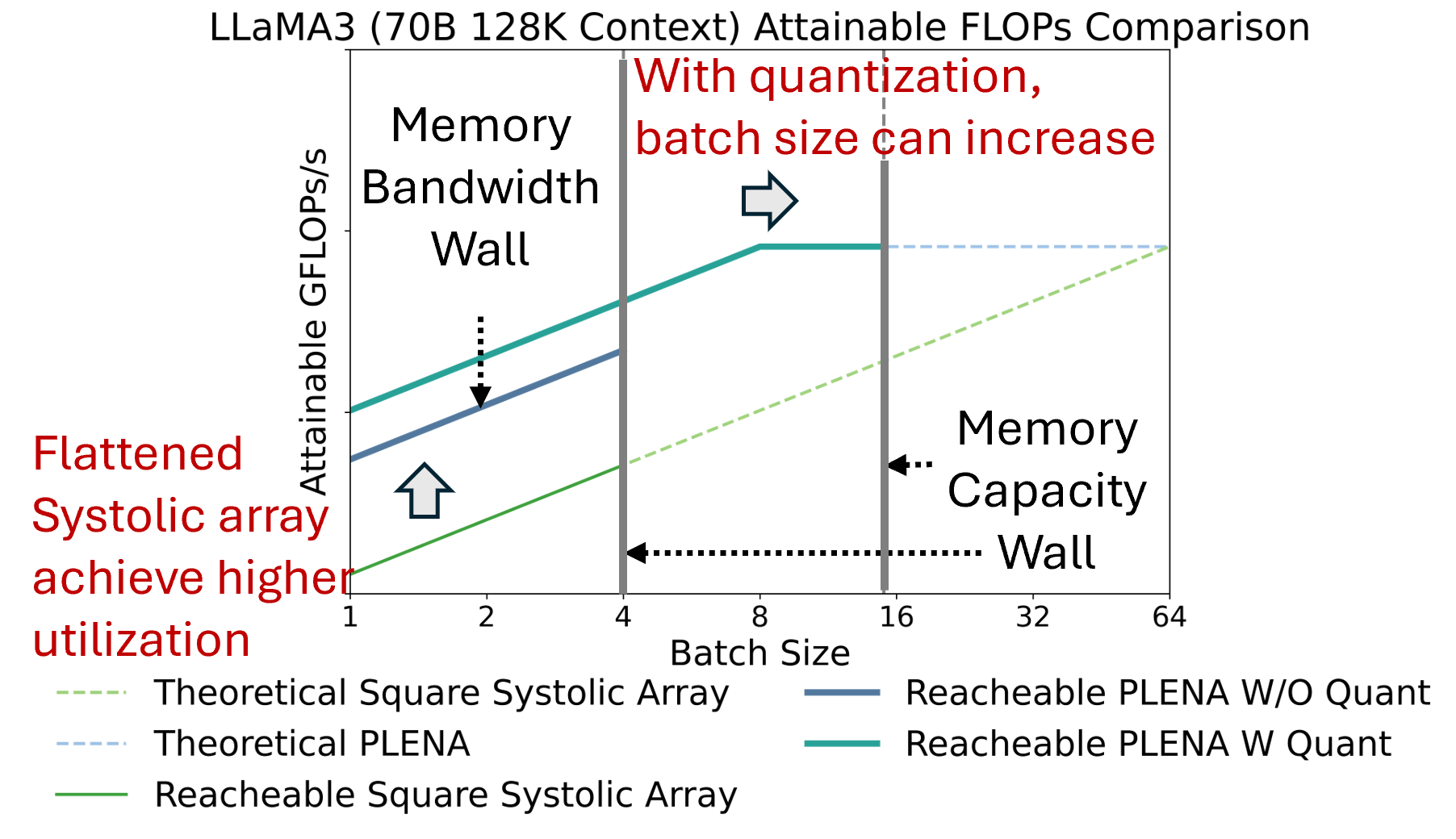}
    \caption{PLENA’s optimization pathways---(1) a flattened systolic array and (2) asymmetric quantization---together achieve improved effective memory bandwidth utilization and help reduce memory capacity limitations.}
    \label{fig:sa_compare_analysis}
  \end{subfigure}

  \caption{A comparison of attainable FLOPs between a square-shaped systolic array (e.g.~TPUs) and PLENA’s when using the same number of multipliers\protect\footnotemark.}
  \label{fig:motivation}
  \vspace{-10pt}
\end{figure}

\footnotetext{64$\times$64 square-shaped systolic array and 8$\times$512 flattened systolic array. Data derived from 144\,GB HBM capacity and 512\,GB/s memory bandwidth.} First, our novel flattened systolic array (\textit{Pathway 1}) resolves the architectural mismatch caused by the typical square-shaped GEMM used for inference, achieving a higher compute utilization of multiplication resources as illustrated in \Cref{fig:sa_compare_plot,fig:sa_compare_analysis}. 
Second, we apply an \textit{asymmetric quantization strategy} with Post-Training Quantization (PTQ) optimizations (\textit{Pathway~2}), where Weights(W)/Activations(A)/KV Cache(KV) can be set to different precisions to alleviate both memory bandwidth and capacity walls. 
With more aggressive W and KV cache quantization, we free up more space in HBM for data scaling (e.g., supporting larger batch sizes). 
\Cref{fig:motivation} shows how these pathways together can \textbf{increase the utilization} compared to the conventional square-shaped GEMM hardware without any optimization.
Finally, as \Cref{fig:Flops_breakdown} shows that attention dominates the compute at longer context lengths, we design PLENA’s custom ISA and novel architecture to effectively support FlashAttention (\textit{Pathway~3})—an IO-aware, fused attention algorithm that substantially reduces off-chip memory traffic~\cite{FlashAttention}. This reduces the likelihood of attention operations saturating memory bandwidth, thereby diminishing the wall’s effect.

Together, these optimization pathways yield significantly higher utilization than conventional square-shaped systolic-array accelerators. The main contributions are as follows:

\begin{itemize}[leftmargin=1em]
    \item We analytically characterize the bandwidth and capacity memory walls in agentic LLM inference and show that existing systolic-array accelerators are normally heavily under-utilized when running agentic workloads.

    \item We introduce three optimization pathways that jointly address the under-utilization caused by memory walls: (i)~a flattened systolic array architecture; (ii)~an asymmetric quantization scheme, coupled with an in-depth exploration of micro-scaling arithmetic's compatibility with optimization techniques such as rotation and norm-guided iterative optimization; and (iii)~a native support for FlashAttention. Together, these enable a holistic approach that addresses both bandwidth and capacity limitations by integrating hardware-level and algorithmic optimizations. 
    
    \item We present PLENA, a complete hardware–software system that realizes the above optimizations. PLENA integrates: (i)~a custom instruction set (PLENA\_ISA) for large Transformer inference; (ii)~a PyTorch-to-PLENA\_ISA compiler; (iii)~an HBM-enabled transactional simulator; (iv)~an automated, accuracy-aware design-space exploration (DSE) flow; and (v)~a full RTL implementation. We demonstrate that PLENA supports different SOTA transformer model variants (e.g., GQA, MHA and MLA~\cite{MLA}, Dense and MoE~\cite{MoE}). We also show that PLENA achieves superior efficiency for agentic LLM inference. Under identical multiplier counts and memory configurations during LLaMA agentic inference, PLENA delivers up to 2.23$\times$ and 4.70$\times$ higher throughput than the A100 GPU and TPU v6e, respectively, and up to 4.04$\times$ higher energy efficiency (Token/J) than the A100. The entire PLENA system will be fully open-sourced upon acceptance.
\end{itemize}

\section{Background and Related Work}

\subsection{Microscaling Data Formats} 
\begin{figure}[h]
  \centering
  \includegraphics[width=1\linewidth]{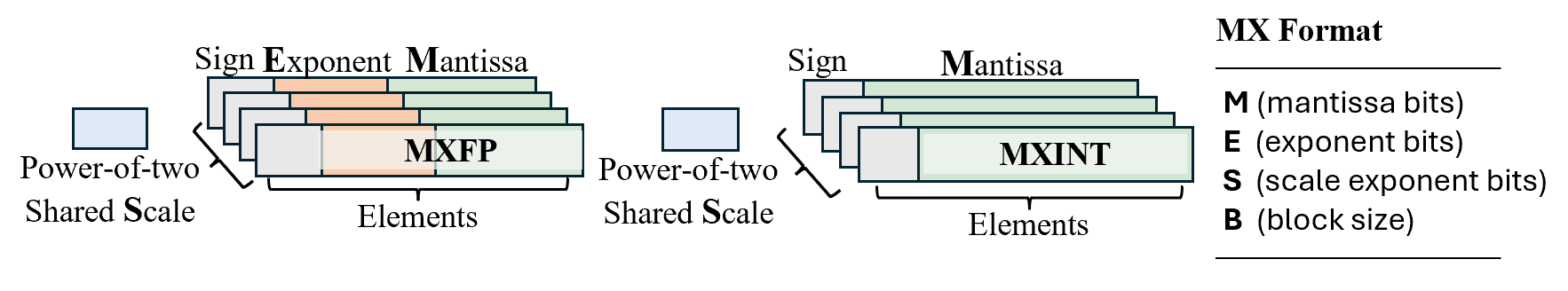}
  \caption{Illustration of the configurable MX data format design, parameterized with tunable configs. Each block of elements shares a power-of-two scaling factor and supports both integer and minifloat data types.}
  \label{fig:MX_Dataformats}
  \vspace{-10pt}
\end{figure}

The concept of block data representation was introduced to collectively represent groups of values using shared scaling factors \cite{darvish2023shared}. 
Building on this idea, Rouhani \emph{et al.}~\cite{rouhani2023microscaling} proposed the Microscaling (MX) data format as a specific variant of block data formats, where each block of elements shares a common scale encoded in an \texttt{E8M0} power-of-two format. MX formats have since been standardized by the Open Compute Project~\cite{rouhani2023ocp}. Recent extensions explore multi-level scaling, where scaling factors are applied hierarchically across granularities. MicroScopiQ~\cite{Microscopiq} adopts a two-level scaling scheme with coarse block-level and finer micro-block-level scales, while NVFP4 \cite{abecassis2025pretraining} employs a similar hierarchy, using a tensor-level \texttt{E8M23} scale and block-level \texttt{E4M3} scale. To balance hardware complexity and software performance, we adopt a single-level scaling scheme in our configurable MX data format, with tunable parameters $(M, E, S, B)$ for MXFP and $(M, S, B)$ for MXINT, illustrated in \Cref{fig:MX_Dataformats}.

\subsection{Co-designing PTQ with Microscaling Data Formats}

Existing off-the-shelf Post-Training Quantization (PTQ) methods are well-studied for integer data formats \cite{frantar2022gptq, QuaRot}. However, we find that these methods are less explored—and in some cases, not directly applicable—to the MX data format. 

GPTQ \cite{frantar2022gptq} was originally developed for integer quantization. We explore its adaptation to our parameterized MX data formats and propose a variant method that better adapts it to the MX format. Details are deferred to~\Cref{subsec:weights_quant}. Rotation-based PTQ methods are among the most effective techniques for mitigating activation outliers. QuaRot~\cite{QuaRot} demonstrated that the application of the Hadamard transformation can effectively suppress such outliers. However, we empirically experimented and found that without careful treatment, the direct application of these methods can lead to significant model performance degradation for MX data formats. Details are deferred to~\Cref{subsec:act_quant}.
Overall, our co-designed PTQ and data format achieves performance competitive with full-precision baselines, even under aggressive low-bit settings, as demonstrated in~\Cref{subsec:quantization_experiments}.

\begin{figure*}[t]
  \centering
  \includegraphics[width=0.96\textwidth]{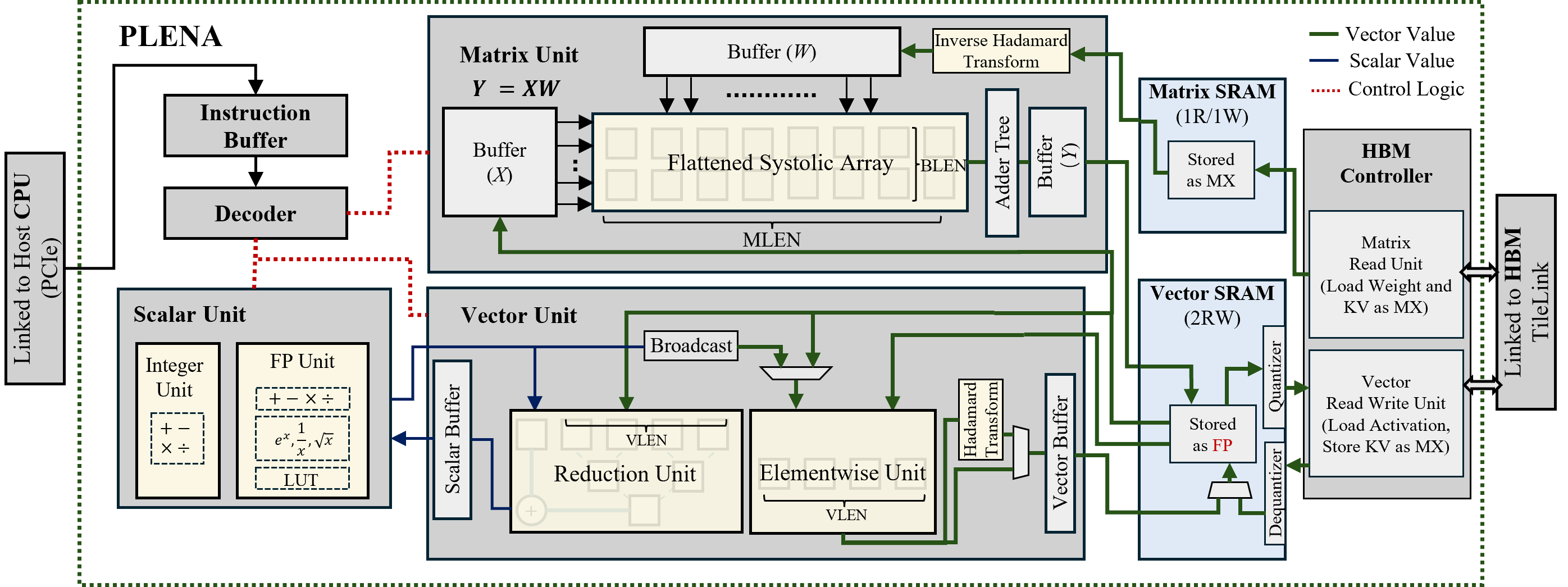}
  \caption{PLENA accelerator architecture overview. Execution is controlled by the decoder’s system-pipeline controller, which derives control signals from decoded instructions and monitors memory dependencies. For example, if the current instruction needs to read from a Vector SRAM row that is still being updated by the vector or matrix unit, the controller inserts a stall to ensure correctness. Vector SRAM acts as the on-chip scratchpad to the matrix and vector units.}
  \label{fig:PLENA_arch}
\end{figure*}

\subsection{FlashAttention}

FlashAttention optimizes memory I/O in the standard attention layer~\cite{FlashAttention}. In a standard attention layer, computing \(QK^{\top}\) produces a prohibitively large square matrix, often thousands by thousands in size. Because on-chip memory cannot hold this intermediate result, it must be written to off-chip memory and later reloaded for the subsequent softmax and \(PV\) steps, which significantly degrades performance. FlashAttention avoids this round trip by tiling and fusing the attention computation (GEMM–Softmax–GEMM) so that all intermediate results fit on-chip. 

Most existing systolic-array–based accelerators do not natively support FlashAttention. SystolicAttention~\cite{SystolicAttn} is among the first to integrate FlashAttention into a systolic architecture by deploying the FlashAttention into the hardware.  In contrast, PLENA adopts a more flexible approach, enabling aggressive memory prefetching overlap and leveraging a mixed-precision supported flattened systolic array with head-level decomposition to achieve higher compute utilization and efficiency. As discussed in \Cref{subsec:flashatten}, we identify three key architectural capabilities required to efficiently support FlashAttention.

\subsection{Accelerators and Their Quantization Supports}
\label{sec:accelerators}

Recent LLM accelerators~\cite{picachu, Microscopiq, FlightLLM, TENDER, FIGNA, Tandem, guo2023olive, hu2025mant, guo2022ant} explore diverse architectural trade-offs across compute organization, kernel specialization, and system integration. However, many of these designs focus on accelerating specific kernels (e.g., GEMM or attention) rather than supporting the full Transformer inference pipeline, often requiring offloading of unsupported operations to external processors. Such partial coverage can introduce additional data movement and limit sustained utilization under long-context inference workloads. PLENA instead targets full Transformer inference directly on the accelerator fabric.

Prior works have also explored hardware and quantization
co-design~\cite{zadeh2020gobo, guo2022ant, guo2023olive, Microscopiq,
hu2025mant}. MicroScopiQ~\cite{Microscopiq} adopts GPTQ for two-level
MX quantization. ANT and MANT~\cite{guo2022ant, hu2025mant} propose
hybrid data formats that adapt quantization mode to input distributions
at runtime. OliVe~\cite{guo2023olive} handles outliers by pairing them
with adjacent low-magnitude weights. However, these
works mostly focus on weight and activation quantization, without
jointly addressing KV cache quantization under long-context inference
scenarios. PLENA, by contrast, is the
first to natively support tunable MX formats with both
hardware friendly QuaRot~\cite{QuaRot} and GPTQ~\cite{frantar2022gptq} while
targeting long-context workloads natively.

Prior work such as Scale-Sim~\cite{samajdar2018scale} supports the simulation of flattened systolic arrays for DNN inference, while SARA~\cite{SARA} explores reconfigurable array shapes to optimize general DNN workloads. However, these approaches do not explicitly consider the characteristics of autoregressive Transformer inference. PLENA instead adopts a workload-driven design that reshapes the systolic organization to address the imbalance between FlashAttention and FFN computation under memory-constrained batching. The flattened array is further optimized for FlashAttention via head-level decomposition, enabling efficient acceleration of both FFN and attention during prefill and decode while maintaining high utilization for both standard and long-context (agentic) inference.

\newcommand{\myflushright}[1]{%
  \unskip\hspace*{1em plus 1fill}%
  \nolinebreak[3]\hspace*{\fill}\mbox{\upshape #1}
}

\newcommand\capped[1]{\textcolor{czblue}{\boldsymbol{#1}}}
\newcommand\varied[1]{\textcolor{czred}{\boldsymbol{#1}}}

\definecolor{nonlin}{RGB}{33,113,181}   
\definecolor{linop}{RGB}{35,139,69}     
\definecolor{matmul}{RGB}{180,15,32}    
\newcommand{\nonlin}[1]{\textcolor{nonlin}{\textsc{#1}}}
\newcommand{\linop}[1]{\textcolor{linop}{\textsc{#1}}}
\newcommand{\mat}[1]{\textcolor{matmul}{\textbf{[MatMul#1]}}}

\section{PLENA Hardware System}

The overall configuration of PLENA is shown in \Cref{fig:PLENA_arch}. It employs instruction-level pipelining and mainly consists of three compute units: the Matrix Unit, the Vector Unit, and the Scalar Unit. All units are highly configurable, supporting multiple data types and precisions (\Cref{tab:design_space}), enabling the application of different quantization methods to the accelerator.

PLENA also includes two main on-chip SRAM blocks. The Vector SRAM acts as a scratchpad for computation, storing frequently used data such as activations, which do not need to be written back to HBM, thereby reducing memory access overhead. The custom Matrix SRAM is dedicated to loading weights and KV tensors and supports reading data in either transposed or untransposed access patterns with minimal extra resource cost and access overhead. 

\subsection{Asymmetric Arithmetic Data Path}
\label{sec:asymmetric}

To support asymmetric quantization strategies, PLENA natively supports multiple numeric formats---covering different data types and precisions---across its compute and memory units. This innovative \emph{asymmetric} data-handling configuration has the following characteristics.

(i)~Activations are stored in a high-precision floating-point (FP) format on-chip in the Vector SRAM, as they are more sensitive to quantization errors than KV or weights. (ii)~KV and weights, being less accuracy-sensitive, can be more aggressively quantized and staged in the Matrix SRAM using lower-precision MX formats (MXFP or MXINT). (iii)~An optional on-chip rotation step can suppress outliers before quantization to preserve accuracy.

Furthermore, when appending new \(K\) and \(V\) vectors to the KV cache in HBM during attention, we selectively apply a Hadamard-based rotation (algorithm detailed in \Cref{subsec:act_quant}) to suppress outliers before quantizing them to the MX data type and storing them in HBM. Since \(K\) and \(V\) are consumed exclusively by the attention GEMMs, they are loaded directly into the Matrix SRAM, where the inverse Hadamard transform is applied before use. These rotation/de-rotation stages can be selectively applied per tensor; for example, weights loaded into the matrix unit bypass the inverse transform.

\subsection{Flattened Systolic Array}

\begin{figure}[!t]
  \centering
  \includegraphics[width=0.48\textwidth]{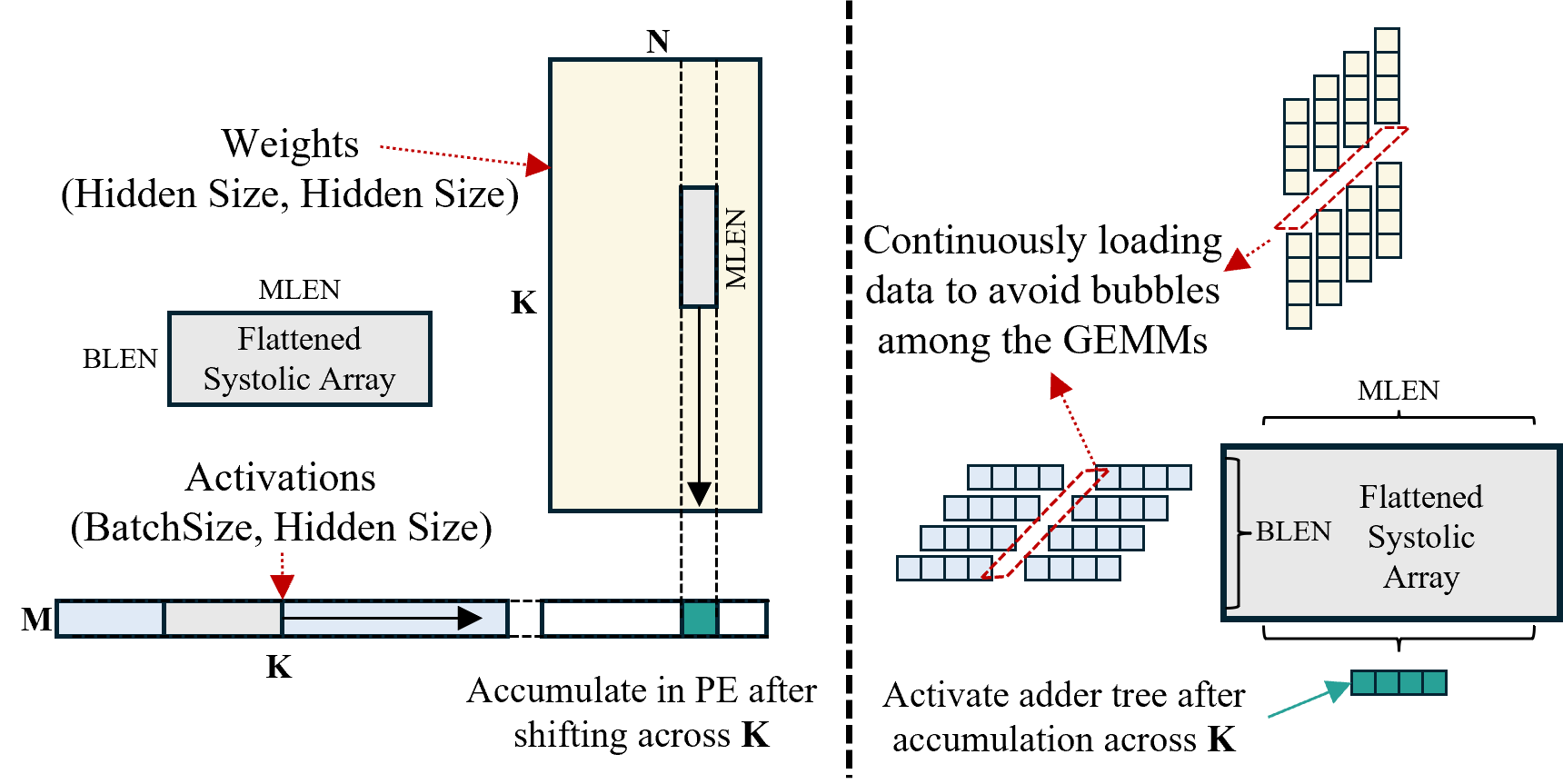}
  \caption{Processing flow for the weight--activation output stationary GEMM. Because memory capacity constrains batch size, the \(M\) dimension remains small. Setting \(\texttt{BLEN}=M\) on the flattened systolic array yields high utilization.}
  \label{fig:SA_Processing Flow}
  \vspace{-10pt}
\end{figure}

As shown in \Cref{fig:sa_compare_analysis}, long-context workloads frequently involve \textit{fat GEMMs} during the feed-forward (FFN) computation, where the batch-related dimension (typically $M$ in $(M,K)\times(K,N)$) is much smaller than the others, resulting in uneven matrix shapes (\Cref{fig:SA_Processing Flow}), while the reduction dimensions \(K\) tend to be very long, for example, the weight--activation GEMM reduces over the model’s hidden size (e.g., \(4{,}096\) for \llamaIIISmall and \(8{,}192\) for \llamaIIIBig). 

Additionally, in the FlashAttention stage, per-head \textit{fat GEMMs} operations are required. The head dimension is typically small (e.g., 128 for \llamaIIIBig), and the Grouped Query Attention (GQA) paradigm requires each key head to be multiplied by multiple query heads simultaneously. This results in low utilization of large-scale systolic arrays when performing per-head GEMMs in FlashAttention, as the computation dimension becomes relatively small.

To improve hardware efficiency in the two most computationally intensive layers, we propose the \textit{flattened systolic arrays} architecture, which achieves a significantly higher utilization for both layers. For the FFN layer, each processing unit performs a \((\texttt{BLEN}, \texttt{MLEN}) \times (\texttt{MLEN}, \texttt{BLEN})\) GEMM, producing an output of shape \((\texttt{BLEN}, \texttt{BLEN})\). Typically, \(\texttt{BLEN}\) is configured to be much smaller than \(\texttt{MLEN}\) to match the workload characteristics of long-context LLM inference. For the FlashAttention module, the systolic array is partitioned into multiple smaller flattened array cores to support per-head GEMM computations, where each core performs a \((\texttt{BLEN}, \texttt{HLEN}) \times (\texttt{HLEN}, \texttt{BLEN})\) GEMM across \((\texttt{MLEN} // \texttt{HLEN})\) heads in parallel.

\begin{figure}[t]
  \centering
  \includegraphics[width=0.48\textwidth]{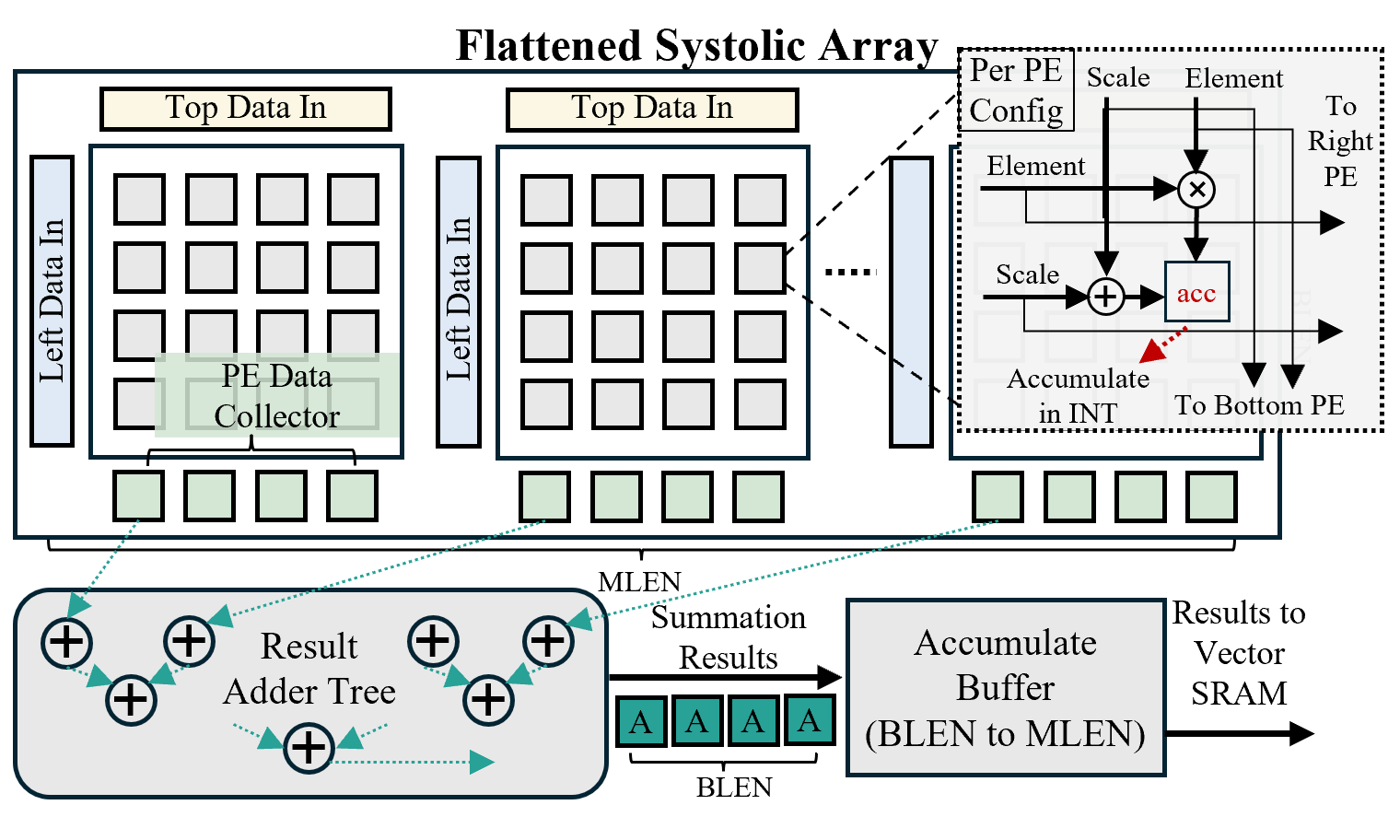}
    \caption{
    At each cycle, the flattened systolic array fetches two \texttt{MLEN}-wide inputs: one from the Matrix SRAM (top) and one from the Vector SRAM (left). The inputs are buffered and reordered, then partitioned into \(\texttt{MLEN}/\texttt{BLEN}\) subvectors (assuming \texttt{MLEN} is divisible by \texttt{BLEN}), each of width \texttt{BLEN}. Each subvector is forwarded to a corresponding sub-array from the top and left directions. 
    The scales and elements are streamed separately to each subarray. For improved resource efficiency, each PE consumes MX-format inputs and performs accumulation in INT precision. The accumulated results are converted to the target activation precision before being written back to the Vector SRAM.
    }
  \label{fig:Flattened_Systolic_Array}
  \vspace{-15pt}
\end{figure}

This flattened systolic array is designed for the output-stationary dataflow in order to maintain a high utilization. As shown in \Cref{fig:SA_Processing Flow}, operands stream along the large reduction dimension \(K\) while partial sums remain stationary in the PEs. The array is then fully pipelined, eliminating idling bubbles between consecutive GEMM tiles.
The microarchitecture of the flattened systolic array is shown in \Cref{fig:Flattened_Systolic_Array}. It is built from a series of small square-shaped systolic arrays (\emph{sub-arrs}), each consisting of a grid of processing elements (PEs). Each PE repeatedly performs multiply–accumulate operations and passes data to its neighboring PEs below and to the right across the array. As described in \Cref{sec:asymmetric}, the systolic array is designed to natively accept data in the MX format.

However, a matrix unit composed solely of \emph{sub-arrs} is insufficient to complete a $(\texttt{BLEN}, \texttt{MLEN})\times(\texttt{MLEN}, \texttt{BLEN})$ GEMM. Each array accumulates only partial sums for a fragment of the result; producing a complete \((\texttt{BLEN}, \texttt{BLEN})\) output requires a cross-array reduction that sums the partial sums held in the PEs across the tiled row. To address this, we integrate a result adder tree (see \Cref{fig:Flattened_Systolic_Array}) that performs the cross-array summation efficiently. This unit is invoked via a dedicated instruction M\_SUM, as only one cross-array summation is required when computing GEMM along the large reduction dimension. This prevents bubbles and improves computational efficiency.

\subsection{Asymmetric Memory Balancing}
Our memory system is characterized by two key properties: 1) Support for asymmetric precisions, variable-length memory transfers, and strided loads/stores to HBM; and 2) Latency hiding for HBM accesses via an memory load unit that operates in parallel with the main execution, enabling high bandwidth utilization.

\begin{figure}[!t]
  \centering
  \includegraphics[width=0.48\textwidth]{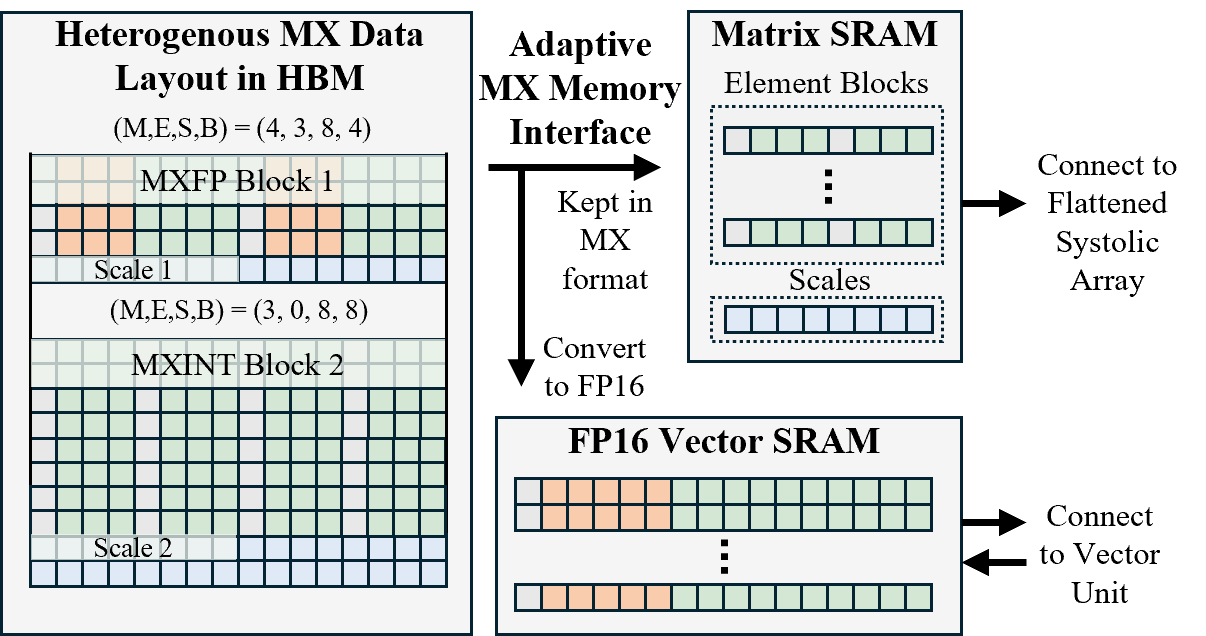}
  \caption{Data layouts and data paths for the memory system. Data with different MX precisions and datatypes are stored following a unified HBM storage pattern. A conversion to FP16 is performed as the data enter the Vector SRAM, which serves as the scratchpad for the vector unit; the vector unit operates in high-precision FP16. For the Matrix SRAM, MX-formatted data loaded from HBM can be stored directly without additional conversion.}
  \label{fig:HBM_sys}
  \vspace{-5pt}
\end{figure}

To make more effective use of HBM capacity, as discussed in \Cref{sec:asymmetric}, all data stored in HBM is kept in the MX format. Since concatenating each data block with its per-block scale would rarely yield a combined size that aligns with a power-of-two memory boundary, we instead store the blocks and their corresponding scales for each tensor separately to ensure that both are properly aligned with the memory boundary. This layout improves memory efficiency while maintaining data locality, as illustrated in \Cref{fig:HBM_sys}.

The memory load unit is critical for fully utilizing HBM bandwidth. Hardware prefetch engines are integrated into both the Matrix and Vector SRAMs, enabling background fetching from HBM and streaming data into each SRAM while the rest of PLENA continues executing other instructions. This sustains full utilization of the matrix unit and avoids stalls due to HBM latency. The two load units are controlled directly by dedicated instructions, H\_LOAD\_M for the Matrix SRAM and H\_LOAD\_V for the Vector SRAM. The load size for each instruction is configurable and specified through the \texttt{M\_Load} and \texttt{V\_Load} parameters.

\subsection{PLENA ISA}

Our customized ISA is designed to cover all operations required for transformer inference. The instructions are structured to balance efficiency with flexibility and are built to support multiple transformer-based models and computation optimizations. In addition to FlashAttention, the ISA also supports different transformer variants, such as MHA, MLA~\cite{MLA}, and MoE~\cite{MoE}. A brief summary is provided in Table~\ref{tab:Custom_ISA}.

\begin{table}[t]
\centering
\setlength{\tabcolsep}{6pt}
\renewcommand{\arraystretch}{1.3}
\caption{An overview of the PLENA ISA.}
\begin{tabular}{
    >{\centering\arraybackslash}m{1.3cm}
    m{5cm}
    >{\centering\arraybackslash}m{1.3cm}
}
\toprule
\textbf{Types} & \textbf{Descriptions} & \textbf{\# Instr.} \\
\midrule
Matrix(M)  & Controls GEMM and GEMV operations, with or without matrix transposition & 6 \\
Vector(V) & Performs elementwise and reduction operations, and rotation for quantization & 13 \\
Scalar(S) & Performs scalar INT and FP arithmetic & 17 \\
HBM(H) & Handles data transfers between HBM and the Matrix/Vector SRAMs & 3 \\
Control(C) & Defines operation settings such as the HBM address, nested-loop configuration, and other execution parameters & 8 \\
\bottomrule
\end{tabular}
\label{tab:Custom_ISA}
\end{table}

\begin{figure}[t]
  \centering
  \includegraphics[width=0.48\textwidth]{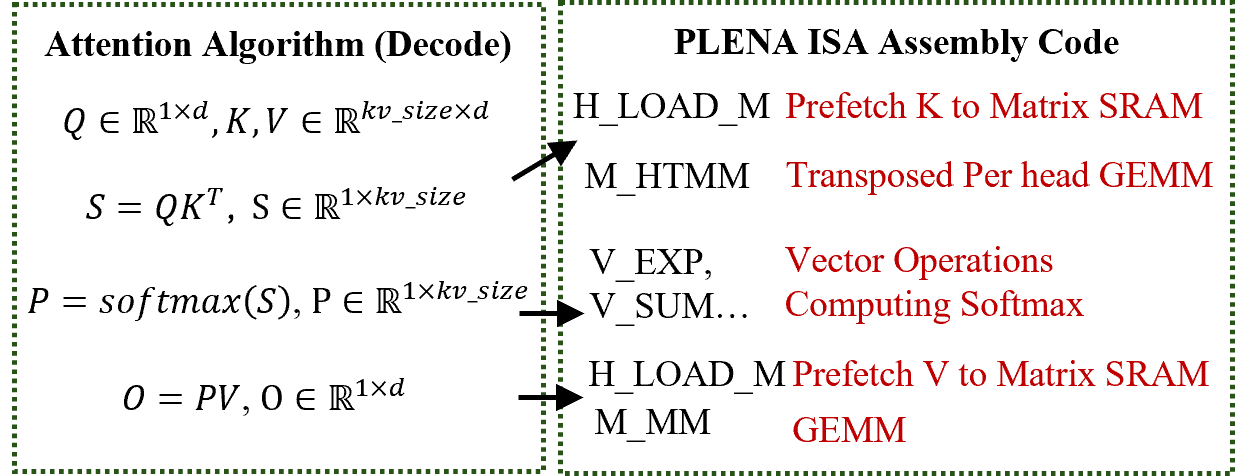}
  \caption{Example of how the single batch single head attention algorithm maps onto PLENA’s custom ISA. Instruction prefixes denote the unit type (e.g., M\_ for Matrix instructions).}
  \label{fig:custom_isa_example}
\end{figure}

To achieve the efficiency and flexibility balance, the ISA is designed to minimize overhead while maximizing utilization of compute and memory resources. This is achieved through features such as tile-level scheduling, which enables fine-grained control of computation and memory instructions at the tile granularity. Furthermore, the ISA defines dedicated instruction classes (Matrix, Vector, Scalar, Memory, and Control) that decouple responsibilities, simplify scheduling, and allow flexible mixing across different computation types.  

Instructions (32 bits each) are dynamically dispatched from the CPU to the instruction buffer via PCIe. In addition to computation, matrix and vector instructions also control read and write operations to their respective SRAMs. Address manipulation is handled by scalar instructions.

\vspace{-2pt}
\subsection{Matrix SRAM}
\label{sec:matrix_sram}

The matrix SRAM is designed to support both transposed and non-transposed accesses without additional latency or data movement overhead. This design specifically targets optimizing the transposed matrix multiplication (\(QK^{\top}\)) in FlashAttention (see \Cref{fig:custom_isa_example}) with low hardware overhead.

In autoregressive inference, explicitly transposing large tiles during the $(QK^{\top})$ computation introduces significant area, energy, and latency overhead. Storing $(K^{\top})$ directly in HBM is also impractical, as newly generated $K$ vectors must be appended to the KV cache during decoding. Consequently, transposition must be performed on the fly, motivating an SRAM organization that supports both row and column access efficiently without explicit data rearrangement.

\begin{figure}[!t]
  \centering
  \includegraphics[width=0.47\textwidth]{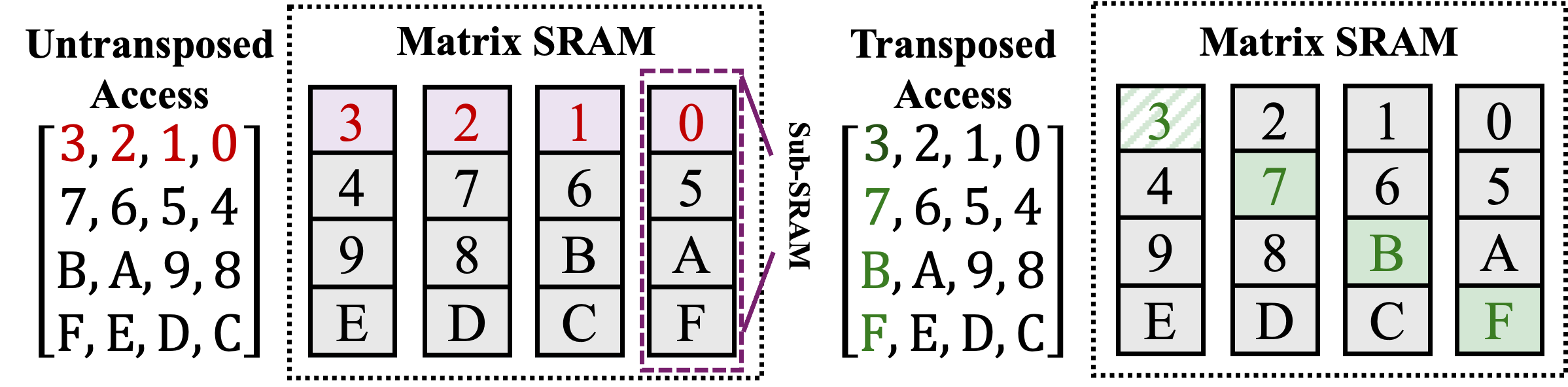}
    \caption{The transposable matrix SRAM design ensures that, for both untransposed and transposed accesses, each sub-SRAM is accessed by at most one element per cycle. As a result, no additional access cycles are introduced.}
  \label{fig:Transpsoed_SRAM}
  \vspace{-10pt}
\end{figure}

As shown in \Cref{fig:Transpsoed_SRAM}, the matrix SRAM distributes each logical row across multiple sub-SRAM banks, storing elements of the same row in different banks at distinct addresses. This layout ensures that row and column accesses map to separate banks, allowing transposed and non-transposed accesses to proceed in parallel without bank conflicts, thereby preserving bandwidth and avoiding explicit data movement.

\subsection{Supporting FlashAttention}
\label{subsec:flashatten}
Most existing systolic-array–based accelerators do not natively support FlashAttention due to these three key elements: 
\begin{enumerate}
    \item They do not support tile-level overlapping of off-chip memory prefetching with computation, resulting in additional latency overhead as execution must wait for data to be loaded from off-chip memory.
    \item They lack memory-layout support such as transpose-on-read and efficient strided/blocked streaming.
    \item They expose only GEMM primitives and lack in-line, row-wise reductions and nonlinear operations (\texttt{max/sum}, \texttt{exp}, \texttt{div}) required for the online softmax.
    \item Their ISAs enforce fixed scheduling and coarse-grained kernel boundaries, which restrict fine-grained tile-by-tile execution and prevent the fused computation pattern.
\end{enumerate}

In PLENA, we address (1) and (2) through the proposed \emph{Matrix SRAM} (see \Cref{sec:matrix_sram}), which enables instruction-level control of memory prefetching and supports transpose-on-read with low overhead. Challenge (3) is addressed by vector and scalar units that implement reductions and element-wise operations. The vector width (\texttt{VLEN}) is configurable to match the tile dimensions used by FlashAttention. The computation precision is also configurable, but is typically set to higher precision (e.g., FP12) to preserve numerical accuracy during the softmax computation. For (4), our custom ISA provides composable, fine-grained control, enabling persistent, tile-by-tile scheduling of the fused attention pipeline. This allows each stage of FlashAttention to be orchestrated individually at tile granularity. Together, these mechanisms enable PLENA to support FlashAttention natively and efficiently.

\begin{figure}[!t]
  \centering
  \includegraphics[width=0.47\textwidth]{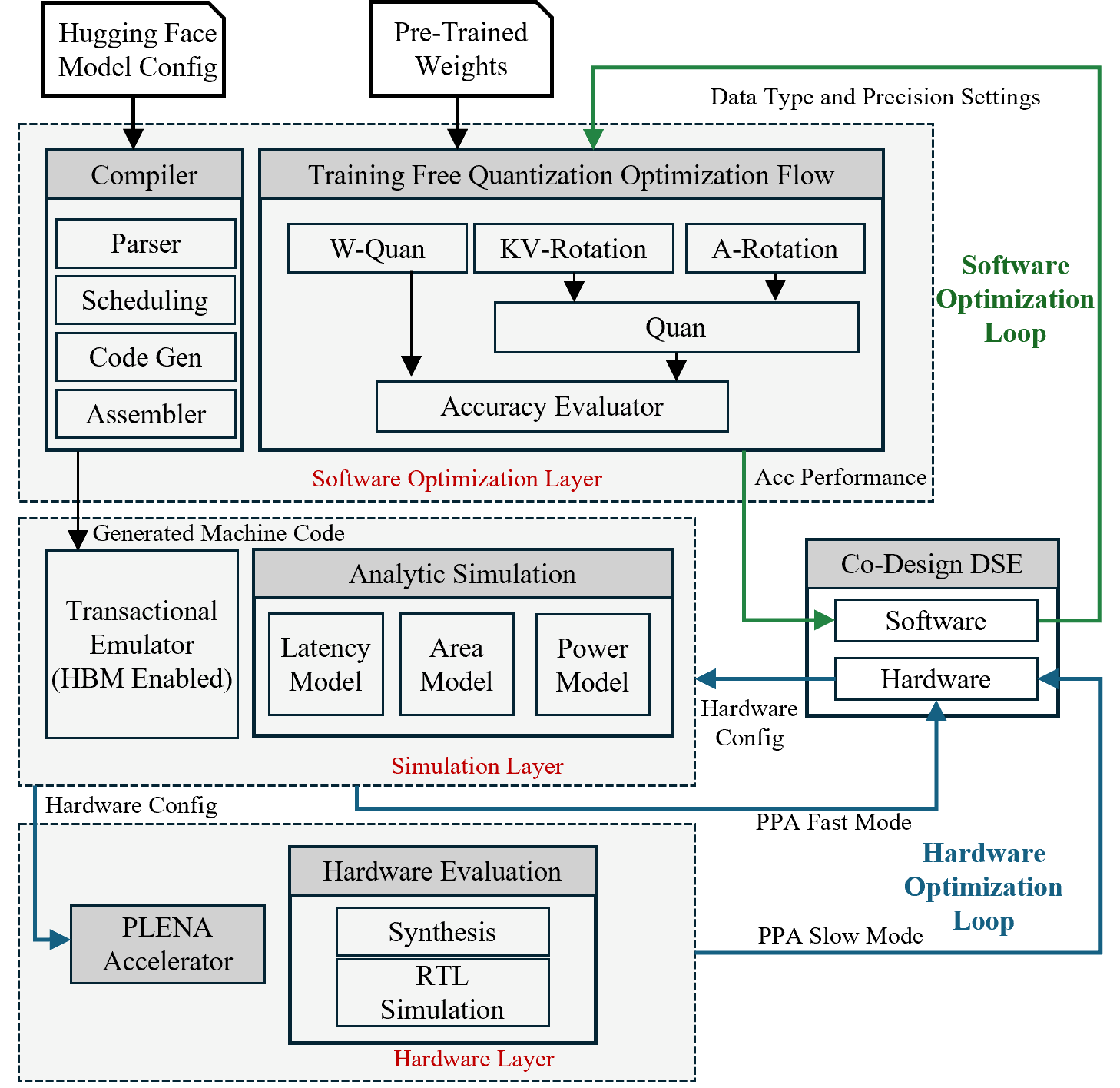}
    \caption{The co-design framework consists of hierarchical layers (actual hardware, transactional emulator, and analytic simulator) with different fidelities. The transaction-level simulator offers good fidelity (cycle-accurate) while achieving an over $200\times$ speedup compared to RTL simulation, and is used for our Co-Design DSE.}
  \label{fig:PLENA_overview}
  \vspace{-10pt}
\end{figure}

\subsection{The PLENA Compilation and Simulation Stack}

PLENA provides a comprehensive design and evaluation framework that can rapidly adapt to new models or new hardware accelerators and optimize for them (\Cref{fig:PLENA_overview}).

Since Transformer computations are highly repetitive and structurally uniform, the PLENA compiler is intentionally kept lightweight: it parses configuration metadata from the model configuration file and maps it onto a predefined PLENA custom ISA assembly template.

To evaluate architectural trade-offs, we developed a transaction-level (cycle-approximate) emulator in Rust that executes the generated machine code in an event-driven manner. The emulator models compute execution, instruction scheduling, and memory transactions at cycle granularity. It is integrated with Ramulator~\cite{Ramulator} and DRAMSys~\cite{dramsys} to provide detailed off-chip memory timing and bandwidth modeling, including bank-level behavior. This enables quantitative analysis of memory–compute interaction, which is critical because memory bandwidth constitutes a primary bottleneck in long-context LLM inference.

The emulator supports the full PLENA architectural design space, including asymmetric mixed-precision arithmetic (\Cref{sec:asymmetric}). By bridging analytic modeling and RTL simulation, it enables accurate evaluation of architectural mechanisms—such as flattened systolic mapping and on-chip FlashAttention, while remaining significantly faster than RTL simulation. We plan to open-source this emulator to facilitate research on LLM accelerator architectures.

\begin{table}[!t]
\centering
\setlength{\tabcolsep}{3pt}
\renewcommand{\arraystretch}{1.15}
\caption{Average error rates across five trials for different simulation levels, compared with RTL and synthesis results for a single Transformer block of the \llamaIIIBig model.}
\begin{tabular}{lcccc}
\toprule
\textbf{Evaluator / Model} & \textbf{Latency} & \textbf{Area} & \textbf{Power} & \textbf{Exe Time} \\
\cmidrule(lr){1-5}
Analytic Simulator       & 11.32\% & 4.79\% & 23.81\% & 8ms \\
Transaction Emulator     & 4.17\% & not supp & not supp & 4.3mins\\
RTL Sim. / Synth.    & ref & ref & ref & 14hrs\\
\bottomrule
\end{tabular}
\label{tab:eval_models}
\vspace{-5pt}
\end{table}

We validated the simulator against our full RTL implementation: it closely matches the RTL synthesis results in both execution latency and numerical accuracy while delivering roughly a $200\times$ speedup, as shown in \Cref{tab:eval_models}.

\section{Quantization}
\label{sec:quantization}

Our work is closely related to prior studies that use the microscaling data format \cite{microscaling, Microscopiq}. Nonetheless, we highlight in our work that while existing SoTA PTQ optimizations -- such as rotation \cite{QuaRot} and norm-guided optimization \cite{frantar2022gptq} are beneficial for integer quantization, they do not align well with the microscaling format.
We identify these caveats for applying PTQ optimization techniques to microscaling arithmetics:

\begin{enumerate}
    \item For weight quantization, MXFP is generally incompatible with these PTQ optimizations. MXINT demonstrates compatibility, but naively applying it leads to degradation. We introduce a novel block-wise clipping optimization that naturally complements block-based arithmetic like MXINT (\Cref{sec:quantization:gptq}).
    \item For activation quantization, rotation schemes such as QuaRot, when naively applied, lead to performance degradation for both MXINT and MXFP. A performance boost is realized only when they are selectively applied to activations (\Cref{subsec:act_quant}).
\end{enumerate}

In summary, we point out that MXINT with PTQ optimization is the de-facto approach for weight quantization. Meanwhile, activation quantization can utilize MXINT or MXFP, but rotation should be applied only selectively. 
The rest of the section elaborates on these optimization strategies and the root causes of incompatibilities, with \Cref{subsec:co_design} detailing the integration of these quantizations into the PLENA system to facilitate a software-hardware co-design.

\subsection{Preliminaries}
We start by formalizing MX quantization under a single-level scaling scheme using three elements: the MX data format ($\tau$), the scale factor ($s$), and the zero point ($z$). The MX data format is defined by a tuple $\tau = (d, b, B)$, where $d$ denotes the datatype, $b$ is its bit-width, and $B$ is the microscaling block size. For example, $\tau = (\texttt{INT}, 4, 16)$ corresponds to an $\texttt{MXINT4}$ format with block size $B=16$, while $\tau = (\texttt{minifloat}, 4, 16)$ corresponds to an $\texttt{MXFP4}$ format with the same block size. In both cases, all values within a block share a single block-wise scale factor $s$ and zero point $z$.

For any data format $\tau$, the set of representable values is bounded to a finite interval, which we denote as:
\begin{equation}
\Omega(\tau) = \{\, x \in \mathbb{R} \mid
\min_{(d,b)} \le x \le \max_{(d,b)} \,\}.
\end{equation}

 the representable range $[\min_{\tau}, \max_{\tau}]$ of integer MX formats (i.e., $d = \texttt{INT}$) is given by:
\begin{equation}
\min_{\tau} = -(2^{b-1}-1),
\qquad
\max_{\tau} = \;\;2^{b-1}-1.
\end{equation}

We partition a high-precision tensor $\mathbf{W}$ into blocks $w \in \mathbb{R}^B$ of size $B$. For each block $w$, the scaling factor is
\begin{equation}
s = \frac{\max |w|}{\max_{\tau}}.
\label{eq:scaling_calculating}
\end{equation}

The zero-point $z$ shifts the range for alignment; we adopt symmetric quantization ($z=0$) throughout and omit it from subsequent expressions. Quantization then maps $w$ into the target format $w_{\tau}$ as:
\begin{equation}
\label{equ:clip}
w_{\tau}
=
\mathrm{clip}\!\left(
\mathrm{RTN}\!\left(\frac{w}{s}\right),\;
\min_{\tau},\, \max_{\tau}
\right),
\end{equation}
where $\mathrm{RTN}(\cdot)$ denotes round-to-nearest projection. The corresponding dequantization operator reconstructs an approximation of the original block:
\begin{equation}
Q\!\left(w;\, s, \tau\right) = s \cdot w_{\tau}.
\end{equation}

\subsection{Optimizing Microscaling Clipping for Weight Quantization}
\label{sec:quantization:gptq}

Existing microscaling arithmetic implementations utilize a static clipping strategy, typically using a fixed value (eg. the maximum value) as clipping threshold for each block (see \Cref{eq:scaling_calculating}). However, a distinct advantage of employing smaller blocks is the opportunity for more granular control over numerical values. Consequently, we introduce \emph{microscaling block-wise clipping}, a technique that provides a conscious balancing between the clipping overflow error and the underflow errors for inliers.

For the same sliced block $w$ expressed in format $\tau$, with representable range $[\min_{\tau}, \max_{\tau}]$ and empirical range $[\min_{w}, \max_{w}]$, we introduce a \emph{clipping parameter} $p \in \mathcal{P} \subset [0.5,\, 0.99]$. This parameter shrinks the effective range to $[p\min_{w},\, p\max_{w}]$.

By sweeping over a discrete set $\mathcal{P}$, we can obtain optimal clipping $p^\star$ for a given block:

\begin{equation}
\label{eq:rowwise-argmin}
p^\star
= \arg\min_{p \in \mathcal{P}}
\left\|
w
- \textsc{Q}(w;\, p, \tau)
\right\|_2^2 .
\end{equation}
Here $\|\cdot\|_2^2$ denotes the squared Euclidean norm.

\label{subsec:weights_quant}
Clipping the empirical range introduces a trade-off between the clipping error and the underflow error. This issue is particularly critical for microscaling-based arithmetic, as the block size is relatively small compared to tensor dimensions. Making an optimal selection of clipping ranges can significantly influence performance; in our experiments, optimized clipping improved perplexity by 5.5\% on \llamaIIISmall{} in 4-bits weights only quantization setting.

We then detail our method, where we integrate our clipping optimization directly into GPTQ's iterative error propagation flow, and introduce a new \emph{output-norm guided} blockwise clipping search that \emph{minimizes the quantization error of the output block rather than the weight block}.
Formally, let $\mathbf{X} \in \mathbb{R}^{M \times K}$ be the inputs, and $\mathbf{W} \in \mathbb{R}^{N \times K}$ be the weights. 
Given a linear layer $\mathbf{Y} = \mathbf{X} \mathbf{W}^\top$, we slice the weights across the $K$ dimension with block size $B$ (e.g., $\text{MLEN}$ in an MX data format $\tau$), yielding block slices $\mathbf{W}_b \in \mathbb{R}^{N \times B}$ to be quantized, and similarly we can have activations across the $K$ dimension $\mathbf{X}_b \in \mathbb{R}^{M \times B}$.
Let $\mathcal{P}$ denote the set of admissible clipping percentiles, and let $\textsc{Q}(\cdot;\,P,\tau)$ denote per-row quantization in data format $\tau$, where $P = (p_1,\dots,p_N) \in \mathcal{P}^N$ is a collection of row-wise clipping percentiles, our new optimization is then uses an outer loop optimization with the hessian information $\mathbf{H}_F$ to iteratively calibrate the weight value ($W_b += \boldsymbol{\delta}_F$, adapted from GPTQ). 
\begin{equation}
\label{eq:dual-optimization-description}
\begin{aligned}
& \boldsymbol{\delta}_F =
-\Big(\mathbf{W}_b - \textsc{Q}\big(\mathbf{W}_{b};\,P_b^\star,\tau\big)\Big)\,
\Big([\mathbf{H}_F^{-1}]_{bb}\Big)^{-1}
(\mathbf{H}_F^{-1})_{:,b}, \\
& \text{where } \mathbf{H}_F = 2\mathbf{X}_F\mathbf{X}_F^\top.
\end{aligned}
\end{equation}

This is combined with a novel inner loop optimization, which is output-norm guided:
\begin{equation}
\quad P_b^\star
= \arg\min_{P_b\in\mathcal{P}^N}
\left\|
\mathbf{X}_b
\Big(
\mathbf{W}_{b}
-
\textsc{Q}(\mathbf{W}_{b};\,P_b,\tau)
\Big)^\top
\right\|^2_2, 
\end{equation}

\subsection{Selectively Rotated Microscaling Data Formats for Activation and KV Quantization}
\label{subsec:act_quant}

Rotation-based optimization, such as QuaRot\cite{QuaRot} tries to smooth the numerical outlier by introducing a rotation matrix, where $\mathbf{X}, \mathbf{W}, \mathbf{H}$ represent the activation, weight, and Hadamard matrix respectively. 
\begin{equation}
l_{rot}(\mathbf{X}) = \textsc{Q}(\mathbf{X}\mathbf{H}) \cdot \textsc{Q}(\mathbf{H}^{-1}\mathbf{W})
\label{eq:quant_with_rotated_weight}
\end{equation}

Surprisingly, we notice that applying the rotation to finer-grained weight quantization (e.g., MXINT with small block sizes) actually increases perplexity. Intuitively, weights have smaller dynamic ranges compared to activations. The rotation may be unnecessary since most weight outliers are already captured by the shared exponents.

We then propose a \textit{selective rotation} strategy for activation quantization:

\begin{equation}
\begin{aligned}
S &= \arg\min_{s\in\mathcal{M}} \sum_{s\in\mathcal{M}} \Delta_{ppl}(l^*_{rot}), \\
l_{rot}^*(\mathbf{X}) &= \textsc{Q}(\mathbf{X}\mathbf{H}) \cdot \mathbf{H}^{-1} \cdot \textsc{Q}(\mathbf{W}),
\end{aligned}
\label{eq:concise-selective-rotation}
\end{equation}

Now $S$ is a set composed of layers from $\mathcal{M}$, and $\Delta_{ppl}(l_{rot}^*)$ reflects the performance improvement due to rotation for each layer $l$. The objective is to minimize the sum of the performance loss across all layers in $\mathcal{M}$ to select the subset to be included in $S$. Another critical difference is that when such rotation is applied to activations, we have to apply a multiplication with $\mathbf{H}^{-1}$ at run-time, and PLENA provides a native hardware support for this operation. 

\subsection{Asymmetric Quantization and Hardware Co-Design}
\label{subsec:co_design}

As discussed earlier, MXINT is the de-facto quantization for weights, whereas we now exposed a search space for using either MXINT or MXFP for \Cref{subsec:act_quant}. Also, we have to consider various precision setups and hardware design parameters (e.g., tile sizes, load/write sizes). 
We then established a co-design framework to conduct such explorations supported by PLENA's multi-fidelity simulators, as shown in \Cref{fig:PLENA_overview}. 
It is worth noting that our co-design can run at different fidilities as illustrated in \Cref{fig:PLENA_overview}, but we choose to run at the transactional-level, unless specified otherwise, for both reasonable speed and good fidelity.
\Cref{tab:design_space} shows the search space and its related constraints. Our search space considers a range of arithmetic types for A/KV, including MXINT and MXFP, as well as different precision configurations. The result can provide an asymmetrically quantized PLENA accelerator design upon completion of the search.

To automate finding the optimal hardware design and quantization parameters, we propose to employ active learning for design space exploration (DSE). We also provide the capability for investigating the trade-offs between optimizing different objectives. For this, we employ multi-objective Bayesian optimization (BO) in BOTorch, which allows exploring the Pareto frontier in an active manner. In our case, the objective function has three components: accuracy, latency, and chip area: $\mathbf{f} = \big[ f_\text{accuracy}(\cdot), f_\text{latency}(\cdot), f_\text{area}(\cdot) \big]$. The exploration method also accounts for constraints by applying rejection sampling to discard invalid or infeasible candidates. This avoids unnecessary, costly objective evaluations and accelerates convergence of the search.
We first conduct experiments on \llamaTiny{} to enable rapid iteration, and then extend our evaluation to \llamaIIISmall{}. 
The results are described in Section~\ref{subsec:co_design}.

\begin{table}[h]
\centering
\setlength{\tabcolsep}{2pt}
\renewcommand{\arraystretch}{1}
\caption{Selected hardware and quantization parameter co-design search space. 
Example constraints include: (1) memory bandwidth constraint 
$\texttt{MLEN} \cdot \texttt{KV\_WIDTH} \leq \textit{MemBandwidth}$; 
(2) $\texttt{MLEN} \bmod \texttt{BLEN} = 0$; 
(3) $\texttt{MLEN} \geq \texttt{HLEN} \geq \texttt{BLEN}$.}
\begin{tabular}{cll}
\toprule
\textbf{Parameter} & \textbf{Description} & \textbf{Search Range} \\
\midrule
\texttt{BLEN} & Tile size of block unit & [2, 4, ..., 64] \\
\texttt{MLEN} & Tile size of Matrix Unit & [2, 4, ..., 1024] \\
\texttt{VLEN} & Tile size of Vector Unit & [2, 4, ..., 1024] \\

\texttt{M\_LOAD} & Matrix SRAM load amount from HBM \\[-1pt]
                    & (num of matrices loaded per instruct) & [2, 4, ..., 256] \\
\texttt{V\_LOAD} & Vector SRAM load amount from HBM \\[-1pt]
                    & (num of vectors loaded per iteration) & [2, 4, ..., 256] \\
\texttt{V\_WRITE} & Vector SRAM write amount to HBM \\[-1pt]
                    & (num of vectors written per iteration) & [2, 4, ..., 256] \\
\midrule
\texttt{ACT\_WIDTH} & Activation precision & MXINT$^{\dagger}$, MXFP$^{\dagger}$ \\
\texttt{KV\_WIDTH}  & Key/Value precision & MXINT$^{\dagger}$, MXFP$^{\dagger}$ \\
\texttt{FP\_SETTING} & Floating-point precision & FP$^{\dagger}$ \\ 
\bottomrule
\end{tabular}
\label{tab:design_space}
\vspace{-10pt}
\end{table}

\begin{table*}[tb]
\centering
\setlength{\tabcolsep}{5pt}
\renewcommand{\arraystretch}{1.15}
\caption{
Multi-objective search results for configurations from a BoTorch run on \llamaIIISmall.
We showcase four representative design points on the Pareto frontier with different perplexity (↓), latency (seconds ↓), area (\(\mu\text{m}^2\) ↓) trade-offs. 
The complete empirical attainment surfaces of the multi-objective search are in \Cref{fig:codesign_eas}.
\best{Best} results are highlighted. 
}
\label{tab:one_frontier}

\resizebox{\textwidth}{!}{%
\begin{tabular}{lccccccccccc}
\toprule
\multicolumn{9}{c}{Parameters} &
\multicolumn{3}{c}{Metrics} \\
\cmidrule(lr){1-9}
\cmidrule(lr){10-12}
\texttt{BLEN} & \texttt{MLEN} & \texttt{VLEN} &
\texttt{M\_LOAD} & \texttt{V\_LOAD} & \texttt{V\_WRITE} &
\texttt{ACT\_WIDTH} & \texttt{KV\_WIDTH} & \texttt{FP\_SETTING} &
Perplexity $\downarrow$ & Lat (s) $\downarrow$ & Area (\(\text{mm}^2\)) $\downarrow$ \\
\midrule
32 & 512  & 128 & 128 &  64 & 256 & MXFP\_E4M3 & MXFP\_E3M4 & FP\_E4M7 &  6.70 & 0.137 & 137.6 \\
32 & 1024 & 1024& 256 & 256 & 128 & MXINT\_8   & MXINT\_4   & FP\_E3M2 &  6.76 & \best{0.116} & 203.4 \\
8  & 128  &  32 & 128 &   8 & 256 & MXFP\_E3M4 & MXFP\_E3M4 & FP\_E5M6 &  \best{6.54} & 0.166 & 26.45 \\
16 & 128  &  16 &   4 &  16 &  64 & MXINT\_8   & MXFP\_E4M3 & FP\_E3M2 &  6.60 & 0.174 & \best{23.64} \\
\bottomrule
\end{tabular}%
}
\vspace{-10pt}
\end{table*}

\section{Evaluation}

\subsection{Experiment Setup}
\paragraph{Models and Datasets} 

We evaluate our quantization framework on popular open-source LLMs, namely LLaMA-2~\cite{llama2} and LLaMA-3~\cite{llama3}, as well as MoE~\cite{MoE} (e.g.~GPT-OSS) and Qwen3 models. Quantization performance is measured in terms of perplexity on the WikiText-2 dataset~\cite{wikitext2}.

The entire quantization process requires approximately 2–20 GPU hours on NVIDIA H100 GPUs. 

\paragraph{Quantization Baselines} 
We compare against several SoTA quantization methods, including software-based approaches targeting GPUs such as GPTQ~\cite{frantar2022gptq}, OmniQuant~\cite{shao2023omniquant}, and QuaRoT~\cite{QuaRot}, as well as approaches used on hardware accelerators such as Atom~\cite{zhao2024atom} and MicroscopiQ~\cite{Microscopiq}.

\paragraph{Accelerator Implementation} PLENA is implemented in SystemVerilog RTL. We perform synthesis using the Synopsys Design Compiler with the 7\,nm OpenROAD predictive PDK~\cite{openroad}. This helps us to generate area and power estimates at a 1\,GHz clock frequency.

\paragraph{Accelerator Baselines} Since our baselines—MicroscopiQ~\cite{Microscopiq}, FIGNA~\cite{FIGNA}, and Olive~\cite{guo2023olive}—are not fully open-sourced or cannot be evaluated under a consistent technology node and toolchain, we re-implemented their core components and integrated them into the PLENA system for a fair inference performance comparison. Additionally, DeepScale~\cite{DeepScale} is used for overall system performance estimation, scaling all designs to the 7\,nm process. Detailed area and power of the core units are evaluated using our own implementations.

\paragraph{Inference Process}
Instead of comparing only with prior accelerator designs, we also evaluate PLENA against high-performance commercial compute platforms, including GPUs (A100 80GB and H100 80GB) and TPUs (v6e-8), to provide a fair and practical comparison. The GPU experiments are conducted in an environment with Ubuntu~22.04, CUDA~12.8, Python~3.11, PyTorch~2.8.0, and vLLM~0.10 V1. The TPU experiments are conducted in an environment with v2-alpha-tpuv6e software.

\subsection{Balancing Area, Latency and Perplexity via Co-design}

\label{subsec:co_design_exp}

\begin{figure}[!t]
  \centering
  \includegraphics[width=0.485\textwidth]{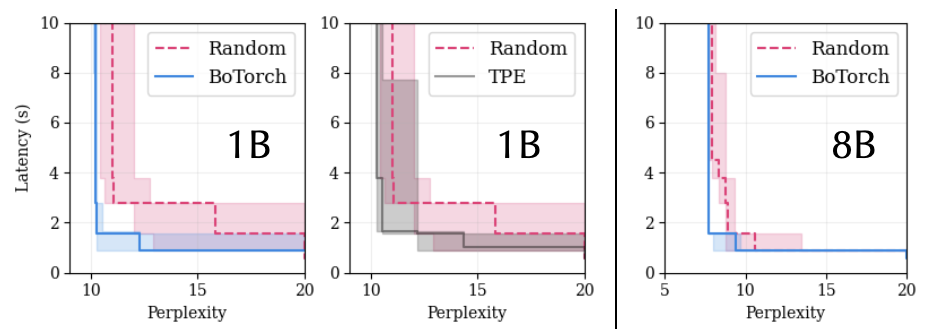}
\caption{Empirical Attainment Surfaces for latency ($\downarrow$) and perplexity ($\downarrow$) objectives across multiple seeds, evaluated with \llamaTiny and \llamaIIISmall over the co-design space shown in \Cref{tab:design_space}.
For the 1B model, we run 9 seeds with 50 trials, comparing BoTorch and TPE methods against Random sampling. 
For the 8B model, we run 5 seeds with 50 trials, comparing BoTorch against Random. 
Shaded regions show the 25\% and 75\% attainment bands across seeds.}
  \label{fig:codesign_eas}
  \vspace{-15pt}
\end{figure}

This subsection shows the results of our design space exploration experiments. Figure~\ref{fig:codesign_eas} shows the Empirical Attainment Surfaces (EAS) for the Pareto fronts found when optimizing with \llamaTiny and \llamaIIISmall. EAS is a visualization approach well-suited for conveying the uncertainty of the Pareto fronts from multiple runs with different random seeds~\cite{knowles2005eas, fonseca2011eas}. Existing tools support visual analysis for two objectives~\cite{watanabe2023eas}, hence we plot EAS for accuracy and latency first. Figure~\ref{fig:codesign_eas} shows that active learning with BoTorch sampler achieves a significantly better tradeoff between latency and perplexity than naive randomized sampling. Tree-Structured Parzen Estimator (TPE) shows more modest gains when optimizing with \llamaTiny compared to using BoTorch sampler, thus we focus on the latter for experiments with \llamaIIISmall.

In \Cref{tab:one_frontier}, we show our co-design results generated from multi-objective optimization runs. These runs can yield designs featuring various trade-offs along the Pareto frontier, with some naturally incorporating multi-precision and multi-arithmetic elements. The PLENA system facilitates such exploration, thanks to its comprehensive simulation and RTL support for these arithmetic types and precision levels.

\subsection{Preserving MX Perplexity via Clipping \& Rotation}
\label{subsec:quantization_experiments}

\paragraph{Main Results}
\begin{table}[t]
    \centering
\caption{\wiki perplexity (\textbf{$\downarrow$}) under GEMM-only emulation (nonlinear ops in full precision) for \llamafamily. 
W/A/KV denote bit widths for weights, activations, and KV cache. Results marked with $^*$ are reproduced from released code.}

\resizebox{1.0\linewidth}{!}{
    \begin{tabular}{lcccccc}
    \toprule
    &   & \multicolumn{3}{c}{\textbf{LLaMA-2} \cite{llama2}} & \multicolumn{2}{c}{\textbf{LLaMA-3} \cite{llama3}} \\
    \cmidrule(lr){3-5}
    \cmidrule(lr){6-7}
    \textbf{Method} & \textbf{W/A/KV}  & \textbf{7B} & \textbf{13B} & \textbf{70B} & \textbf{8B} & \textbf{70B} \\
    \midrule
    Baseline & 16/16/16 & 5.47 & 4.83 & 3.31 & 6.13 & 2.85 \\
    \midrule
    GPTQ \cite{frantar2022gptq} & 4/16/16 & 6.23 & 5.58 & 4.28 & 8.12 & 3.75 \\
    AWQ \cite{lin2024awq} & 4/16/16 & 5.82 & 5.19 & 4.08 & 7.96 & 3.58 \\
    OmniQuant \cite{shao2023omniquant} & 4/16/16 & 5.74 & 5.02 & 3.47 & 7.09 & 3.46 \\
    MicroScopiQ \cite{Microscopiq} & 4/16/16 & 5.65 & 5.02 & 3.42 & 6.89 & \best{3.25} \\
    QuaRot \cite{QuaRot} & 4/16/16 & \best{5.60} & 5.00 &  3.41& 6.52$^*$ &  3.53$^*$ \\
    \textbf{PLENA (MXFP)} & 4/16/16 & 7.09 & 5.91 & - & 11.95 & - \\
    \textbf{PLENA (ours)} & 4/16/16 & 5.61 & \best{4.97} & \best{3.41}& \best{6.45} & 3.59 \\
    \midrule
    OmniQuant \cite{shao2023omniquant} & 4/4/16 & 11.47 & 8.32 & 5.41 & 10.21 & 5.30 \\
    SmoothQuant \cite{xiao2023smoothquant} & 4/4/16 & 20.47 & 15.63 & 17.62 & 29.54 & 19.32 \\
    Atom \cite{zhao2024atom} &  4/4/16 & 6.16 & 6.12 & 5.20 & 8.12 & 4.69 \\
    MicroScopiQ \cite{Microscopiq} &  4/4/16 & 6.11 & 5.57 & 4.48 & 8.12 & 4.65 \\
    QuaRot \cite{QuaRot}& 4/4/16 & 6.02$^*$  & 5.36$^*$ & 3.78& 8.00$^*$& 6.33$^*$ \\
    M-ANT \cite{hu2025mant}& 4/4/16 & 5.92  & 5.24 & - & - & - \\
    \textbf{PLENA (MXFP)} & 4/4/16 & 15.89 & 10.30 & - & 91.71 & - \\
    \textbf{PLENA (ours)} &  4/4/16 & \best{5.69} & \best{5.03} & \best{3.59}& \best{6.76} & \best{4.51} \\
    \midrule
    QuaRot \cite{QuaRot} & 4/4/4 & 6.10 & 5.40 & 3.79&  8.16 & 6.66 \\
    QuaRot-128G \cite{QuaRot}& 4/4/4 & 5.93& 5.26 & \best{3.61} & 7.36&  5.51 \\
    \textbf{PLENA (MXFP)} & 4/4/4 & 67.35 & 27.44 & - & 256.22 & - \\
    \textbf{PLENA (ours)} &  4/4/4 & \best{5.89} & \best{5.18} & 3.62 & \best{7.22} & \best{4.77} \\
    \bottomrule
    \end{tabular}}
    \label{tab:llm_results}
\end{table}

\begin{table}[t]
\centering
\caption{Ablation study of quantization techniques and their impact on microscaling data formats, evaluated across all 9 GEMMs in \llamaIIISmall. Results are reported on \wiki{} perplexity. GPTQ is used for clipping: $\mathrm{Err}_{y}$ denotes output-norm clipping; $\mathrm{Err}_{w}$ denotes weight-norm clipping.}
\label{tab:ablation}

\resizebox{\linewidth}{!}{

\begin{tabular}{lrlr}
\toprule
\textbf{Method} & \multicolumn{1}{l}{\textbf{PPL\textbf{$\downarrow$}}} & \textbf{Method} & \multicolumn{1}{l}{\textbf{PPL\textbf{$\downarrow$}}} \\
\cmidrule(lr){1-2}
\cmidrule(lr){3-4}
\textbf{Baseline FP16} & 6.13 & \multicolumn{2}{l}{\textbf{ACT and KV Only}} \\
\cmidrule(lr){1-2}
\multicolumn{2}{l}{\textbf{Weight Only}} & MXFP4 & 29.75 \\
MXINT + RTN & 6.83 & MXINT4 & 7.24 \\
MXFP + RTN & 11.94 & MXFP4 + Selective Rotate & 14.50 \\
MXINT4 + Rotation & 6.98 & MXINT4 + Selective Rotate & \best{7.05} \\
\cmidrule(lr){3-4}
MXFP4 + Rotation & 13.71 & \multicolumn{2}{l}{\textbf{MXINT Full System}} \\
MXINT4 + $\mathrm{Err}_{w}$ Clip & 6.53 & RTN & 8.28 \\
MXINT4 + $\mathrm{Err}_{y}$ Clip & \best{6.45} & $\mathrm{Err}_{y}$ Clip & 7.60 \\
\multicolumn{2}{l}{\textbf{}} & $\mathrm{Err}_{y}$ Clip +Selective Rotation & \best{7.22} \\
\bottomrule
\end{tabular}
}
\vspace{-13pt}
\end{table}

\begin{table}[t]
    \centering
    \footnotesize
    \setlength{\tabcolsep}{3.5pt}
    \caption{Zero-shot downstream task accuracy of LLaMA-3 models with 4-bit weight, activation, and KV quantization, evaluated on PIQA (PQ), WinoGrande (WG), HellaSwag (HS), Arc-Easy (A-e), Arc-Challenge (A-c), and LAMBADA (LA).}
    \begin{tabular}{l|cccccc|c}
        \toprule
        \textbf{Method} & \textbf{PQ} & \textbf{WG} & \textbf{HS} & \textbf{A-e} & \textbf{A-c} & \textbf{LA} & \textbf{Avg.} \\
        \midrule
        \multicolumn{8}{l}{\textit{LLaMA-3-8B}} \\
        \quad FP16                 & 80.74 & 72.77 & 79.06 & 77.82 & 53.33 & 75.63 & 73.22 \\
        \quad QuaRot~\cite{QuaRot} & 75.14 & 65.82 & 72.94 & 68.01 & 43.34 & 65.81 & 65.18 \\
        \quad \textbf{PLENA (ours)}&79.00  & 71.90 & 76.07 & 74.71 &  48.21 & 72.44 & 70.39 \\
        \midrule
        \multicolumn{8}{l}{\textit{LLaMA-3-70B}} \\
        \quad   FP16   & 84.66 & 80.51 & 84.89 & 85.86 & 64.25 & 79.47 & 79.94 \\
        \quad  QuaRot & 78.07 & 69.30 & 77.33 & 73.44 & 47.53 & 69.57 & 69.21 \\
        \quad  \textbf{PLENA (ours)} &   82.48 &  77.51 &  83.47  & 78.54   &  57.17   &  78.05   &   76.20   \\
        \bottomrule
    \end{tabular}
    \label{tab:zero_shot_combined}
\end{table}

\begin{table}[t]
\centering
\caption{Evaluation of long-context and agentic workloads across code generation (HumanEval~\cite{chen2021codex}), mathematical reasoning (GSM8K~\cite{cobbe2021training}), and function-calling (BFCL-Web Search Base~\cite{patil2025berkeley}) benchmarks. W/A/KV denotes bit widths for weights, activations, and KV cache.}
\label{tab:long_context_agentic}
\begin{tabular}{lc ccc}
    \toprule
    \textbf{Model / Method} & \textbf{W/A/KV} & \textbf{HumanEval} & \textbf{GSM8K} & \textbf{BFCL-W} \\
     & & pass@1 $\uparrow$ & EM $\uparrow$ & Acc $\uparrow$ \\
    \midrule
    \quad Baseline              & 16/16/16 & 89.6 & 97.85 & 27.0 \\
    \quad \textbf{PLENA (ours)} & 4/4/4    & 84.1   & 97.85  & 24.0 \\
    \bottomrule
\end{tabular}
\vspace{-10pt}
\end{table}

\begin{table}[t]
\centering
\caption{Ablation study of quantization techniqeus on Qwen3 model across HumanEval, GSM8K, and BFCL-Web Search Base. $\mathrm{Err}_{y}$ denotes output-norm clipping. Qwen3-8B on HumanEval and GSM8K, Qwen3-32B on BFCL-Web Search Base. Results for HumanEval and GSM8K use Qwen3-8B; results for BFCL-Web Search Base use Qwen3-32B, as its higher baseline accuracy better isolates the effect of quantization.}
\label{tab:ablation_new}
\resizebox{\linewidth}{!}{
\begin{tabular}{l ccc}
    \toprule
    \textbf{Configuration} & \textbf{HumanEval} & \textbf{GSM8K} & \textbf{BFCL-W} \\
     & pass@1 $\uparrow$ & EM $\uparrow$ & Acc $\uparrow$ \\
    \midrule
    Baseline (FP16) & 84.8 & 90.9 & 27 \\
    \midrule
    W-only INT4 (RTN)      & 82.9 & 88.7 & 22 \\
    + ACT \& KV INT4       & 72.0 & 74.4 & 15 \\
    + GPTQ                 & 73.2 & 87.7 & 24 \\
    + $\mathrm{Err}_{y}$ Clip & 74.4 & 88.6 & 24 \\
    + Selective Rotation   & \best{78.7} & \best{88.8} & \best{24} \\
    \bottomrule
\end{tabular}}
\vspace{-10pt}
\end{table}

We evaluate our quantization method against related work; results are summarized in \Cref{tab:llm_results}. For a fair comparison, we first match prior settings by quantizing only the nine GEMMs in the decoder. In the W4A4KV16 setting, our results outperform all related work. For \llamaIIISmall, compared to prior approaches, our method achieves at least a $1.24 \times$ reduction in perplexity. 
We also evaluated the result with the quantized vector cores. 
We find that quantizing the remaining operators to a MiniFloat E6M5 format is effectively lossless in perplexity while reducing memory footprint by 25\% relative to FP16.
The key contributions to this performance improvement come from two aspects:  
1) \textbf{Output-norm guided blockwise clipping search}: by integrating \emph{output-norm guided}, blockwise clipping into iterative weight quantization, we validate that output reconstruction error correlates strongly with end-task performance; consequently, our approach substantially reduces perplexity degradation.
2) \textbf{Selective rotation}: Our approach searches for the best layer-wise rotation combination for each model. Unlike QuaRoT~\cite{QuaRot}, which merges rotation into weights, we apply online rotation only to specific layers. 

\begin{figure*}[h]
  \centering
  \includegraphics[width=1\textwidth]{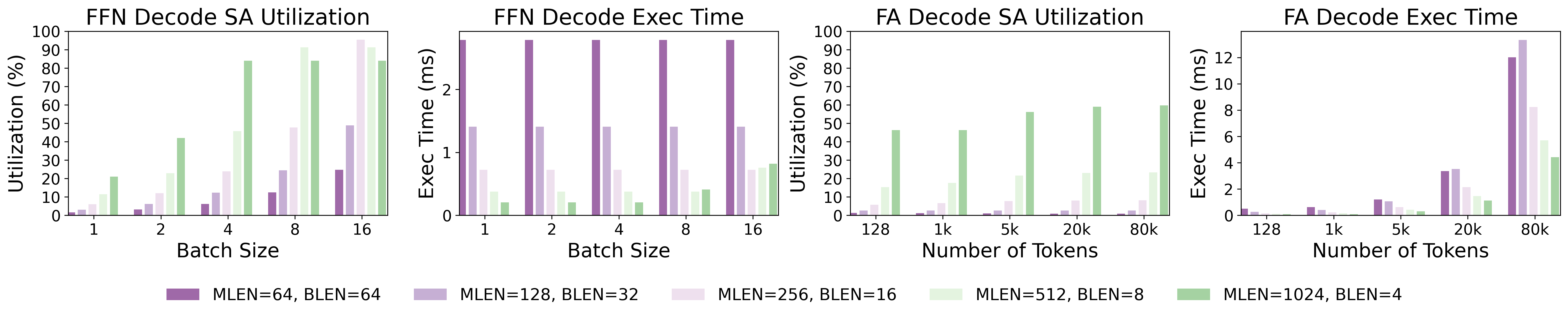}
\caption{The systolic array reaches optimal utilization in the FFN layer when its block length (\texttt{BLEN}) aligns with the batch size.  
FA  = Flash Attention. SA = systolic array.
For FA, flattening the array enhances utilization by allowing parallel processing of multiple attention heads, and is particularly efficient for long-context inference with smaller effective batch sizes.}
  \label{tab:decode_FFN_SA}
\vspace{-10pt}
\end{figure*}

\begin{figure*}[t]
  \centering
  \includegraphics[width=1\textwidth]{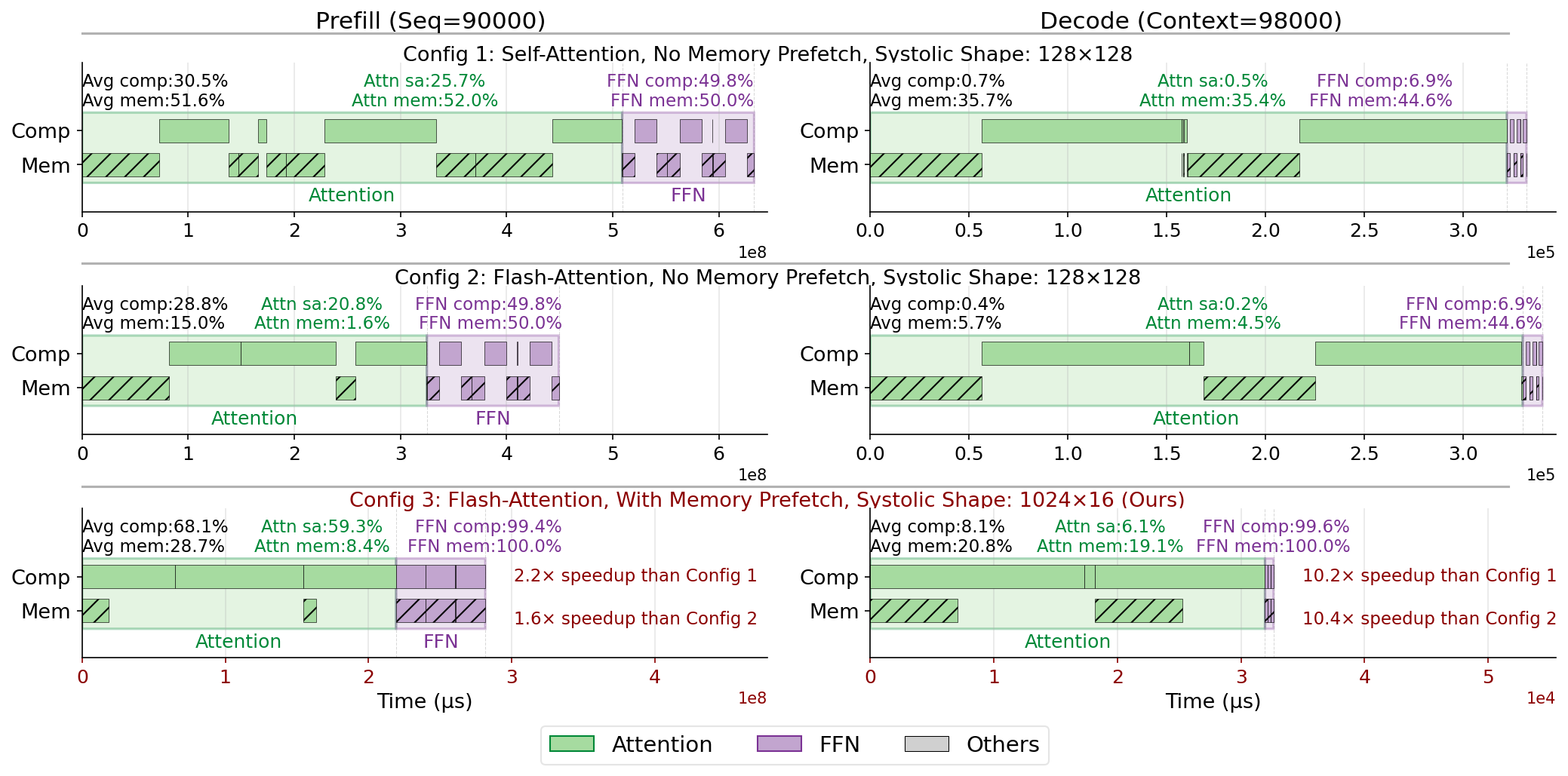}
  \caption{This figure shows the timing performance breakdown of PLENA across the prefill and decode stages for the \llamaIIIPLUSBIG model with a batch size of 16. The breakdown includes compute active time (Comp), memory active time (Mem), systolic array (SA) utilization, and memory bandwidth utilization across the overall inference flow. With on-chip FlashAttention support, large intermediate activations are retained on-chip rather than written to off-chip memory, substantially reducing memory traffic, while memory prefetching hides most data-access latency. In addition, the flattened systolic array configuration maintains high utilization across both prefill and decode stages. For attention workloads, the flattened array achieves high compute and memory utilization by enabling parallel multi-head execution and head preloading.}
  \label{fig:hardware_ablation_study}
  \vspace{-5pt}
\end{figure*}

\paragraph{Abalation Study}

To further assess the impact of our key contributions, we conduct an ablation study on \llamaIIISmall to validate the effectiveness of 1) Output-norm guided blockwise clipping search and 2) Selective rotation. The ablation is structured into three stages: (i) \emph{weight-only} quantization, (ii) \emph{activation \& KV-cache} quantization on top of quantized weights, and (iii) \emph{full-system} emulation where all MX-aware operators are quantized. We show them in \Cref{tab:ablation}. 

First, MXFP4 always underperforms MXINT4 in all settings. Motivated by this, we adopt MXINT as the default data type for all subsequent evaluations. 
Second, for weight-only quantization, we show that rotation generally hurts performance -- they are simply not compatible with microscaling arithmetic. Furthermore, we demonstrate that our output-norm guided block-wise clipping ($\mathrm{Err}_{y}$) achieves better performance compared to weight-error guided block-wise clipping ($\mathrm{Err}_{w}$). 
Third, selective rotation effectively enhances activation and KV quantization for both MXFP4 and MXINT4. This is different from our observations with weight quantization, where rotation negatively impacts perplexity. We hypothesize this arises from the broader numerical range found in activation and KV values, which benefits from rotation's ability to temper the presence of outliers.
Finally, our full-system results confirm that both \emph{block-wise clipping search} and \emph{selective activation rotation} improve the overall performance.

\begin{table}[t]
\centering
\caption{Impact of quantization configurations on memory footprint and bandwidth
for \llamaIIIPLUSBIG under the OSWorld-L workload (90k prefill, 8k output tokens) in \Cref{tab:benchmark_token_usage} with batch size $B=8$.
W/A/KV denote the bit precision of weights, activations, and KV cache, respectively.}
\label{tab:quantization_levels}
\setlength{\tabcolsep}{5pt}
\renewcommand{\arraystretch}{1.05}
\begin{tabular}{lcccc}
\toprule
\textbf{W/A/KV (bits)} & 16/16/16 & 4/16/16 & 4/4/16 & 4/4/4 \\
\midrule
Peak Bandwidth (GB/s) & 8192 & 8192 & 5120 & 2048 \\
KV Cache Footprint (GB) & 239.26 & 239.26 & 239.26 & 59.81 \\
Weight Storage (GB) & 129.46 & 32.36 & 32.36 & 32.36 \\
\bottomrule
\end{tabular}
\vspace{-5pt}
\end{table}

\subsection{Improving Utilization via Flattened Systolic Arrays}

The utilization analysis of PLENA’s flattened systolic array for the FFN and FlashAttention (FA) layers of the \llamaIIISmall model is summarized in \Cref{tab:decode_FFN_SA}. Results for the prefilling stage are omitted because both FFN and FA operate at near-maximum utilization during this phase. For FlashAttention, the computation pattern is independent of batch size, so it is not included in this table. For FFN, the computation for FFN is less important with the growth of generated token length, hence not included as well.

\begin{figure}[!t]
  \centering
  \label{fig:systolic_shape_comparison}
  \includegraphics[width=0.485\textwidth]{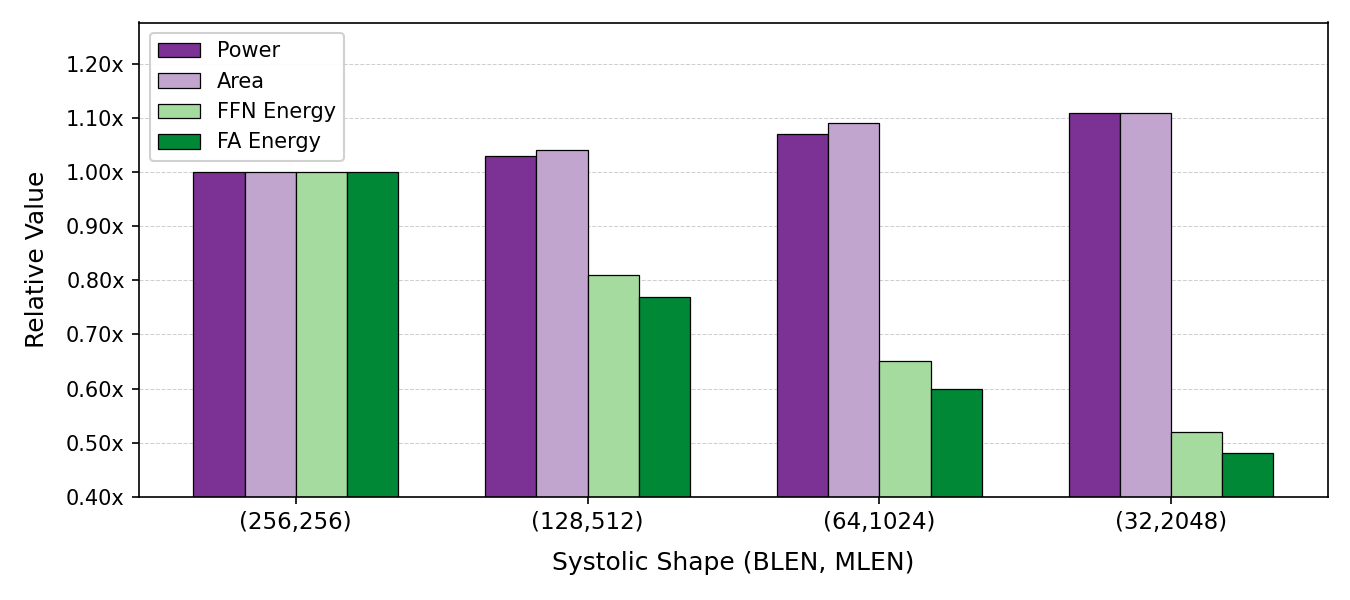}
    \caption{Power and area comparison of matrix units with different systolic array shapes. Although the flattened systolic array incurs slightly higher area and power, its higher utilization leads to significantly lower effective energy consumption for FFN and attention workloads in the agentic task OSWorld-L.}
    \vspace{-10pt}
\end{figure}

The DC synthesis results are reported in \Cref{tab:sa_hardware_comparison}. These results show that the flattened systolic array achieves higher compute-resource utilization for both the FFN and FlashAttention layers compared with prior accelerators. Furthermore, \Cref{fig:systolic_shape_comparison} demonstrates that the flattened systolic organization provides higher energy efficiency, despite some power and area overhead compared with conventional square arrays.

The overall ablation study of the systolic-array optimizations is presented in \Cref{fig:hardware_ablation_study}. The results show that the flattened systolic array, combined with native FlashAttention support, significantly reduces the execution time of both attention and FFN components across the prefill and decode phases, particularly for long-context inference.

\begin{table}[t]
  \centering
  \caption{Compute area, utilization, and attainable FLOPs for systolic arrays. 
    Baselines use \(64\times64\); PLENA uses \(4\times1024\). 
    \emph{S.A = Standard Attainable FLOPs in GSM8K (1.4k/200); A.A = Agentic Attainable FLOPs in OSWorld-L workload (90k/8k) in \Cref{tab:benchmark_token_usage}}}
  \label{tab:sa_hardware_comparison}
  
  \resizebox{1\linewidth}{!}{
  \begin{tabular}{lcccc}
    \toprule
    Metric & MicroscopiQ~\cite{Microscopiq} & Olive~\cite{guo2023olive} & FIGNA~\cite{FIGNA} & \textbf{PLENA} \\
    \midrule
    Comp Area (\(\mathrm{mm}^2\))   & 0.1378 & 0.319  & 0.471  & 0.237 \\
    TOPs/\(\mathrm{mm}^2\)          & 59.45  & 25.66  & 17.39  & 34.49 \\
    S.A FLOPs/\(\mathrm{mm}^2\)$^*$ & 3.36  & 1.60  & 7.51   & \best{29.31} \\
    A.A FLOPs/\(\mathrm{mm}^2\)$^*$ & 1.08   & 0.40   & 6.71   & \best{12.81} \\
    \bottomrule
  \end{tabular}
  }
  \vspace{-15pt}
\end{table}

\begin{table*}[t]
  \centering
  \captionsetup{skip=2pt}
  \caption{System-level comparison across workloads in \Cref{tab:benchmark_token_usage}. Performance evaluation occurs under full HBM-capacity utilization, setting the batch size (BS) to the largest fitting value per workload-hardware pair. Note: We reproduced MicroScopiQ~\cite{Microscopiq} and deployed its compute unit on the PLENA platform for testing. And for GPT-OSS 20B (MoE)~\cite{MoE} and Qwen3-32B~\cite{qwen3}, the remaining accelerators and TPUs are not included since they do not support these configurations~\cite{vllm_tpu_supported_models}.}
  \label{tab:system_level_combined_reformat}
  \setlength{\tabcolsep}{3pt}
  \renewcommand{\arraystretch}{0.9}
  \scriptsize
  \begin{tabular}{lcccccccccccccccc}
    \toprule

\multicolumn{17}{c}{\textbf{\llamaIIIOneSMALL}} \\
\addlinespace[2pt]
& \multicolumn{4}{c}{\textbf{(1.4k, 0.2k)}}
& \multicolumn{4}{c}{\textbf{(114k, 5k)}}
& \multicolumn{4}{c}{\textbf{(90k, 8k)}}
& \multicolumn{4}{c}{\textbf{(90k, 8k) Equal Batch}} \\
\cmidrule(lr){2-5}\cmidrule(lr){6-9}\cmidrule(lr){10-13}\cmidrule(lr){14-17}
\textbf{System}
& TTFT (s) & TPS (×A100) & Tok/J & BS
& TTFT (s) & TPS (×A100) & Tok/J & BS
& TTFT (s) & TPS (×A100) & Tok/J & BS
& TTFT (s) & TPS (×A100) & Tok/J & BS \\
\midrule
A100
  & 0.68 & 1.00x  & 1.00x  & 2048
  & 7.40 & 1.00x  & 1.00x  & 16
  & 5.00 & 1.00x  & 1.00x  & 16
  & 5.00 & 1.00x  & 1.00x  & 16 \\
A100 QuaRot~\cite{QuaRot}
  & 0.73 & 1.12x & 1.12x & 4096
  & 8.63 & 1.10x & 1.10x & 32
  & 5.97 & 1.14x & 1.14x & 32
  & 4.79 & 1.08x & 1.08x & 16 \\
H100
  & 2.42 & 1.65x & 0.94x & 2048
  & 2.66 & 2.50x & 1.43x & 16
  & 1.83 & 2.48x & 1.41x & 16
  & 1.83 & 2.48x & 1.41x & 16\\
H100 QuaRot~\cite{QuaRot}
  & 2.51 & 1.77x & 1.01x & 4096
  & 2.97 & 2.57x & 1.47x & 32
  & 2.01 & 2.55x & 1.46x & 32
  & 1.77 & 2.51x & 1.43x & 16 \\
TPU v6e
  & 5.61 & 0.88x & N/A & 2048
  & 7.58 & 0.51x & N/A & 16
  & 7.23 & 0.53x & N/A & 16
  & 7.23 & 0.53x & N/A & 16 \\
MicroScopiQ~\cite{Microscopiq}
  & 3.47 & 0.83x & 1.67x & 8192
  & 21.28 & 0.37x & 0.74x & 64
  & 19.13 & 0.39x & 0.78x & 64
  & 4.93 & 0.27x & 0.54x & 16 \\ 
\textbf{PLENA}
  & 3.41 & \best{1.91x} & \best{3.50x} & 8192
  & 20.13 & 1.45x & \best{2.66x} & 64
  & 18.87 & 1.45x & \best{2.65x} & 64
  & 4.68 & 1.17x & \best{2.10x} & 16 \\
\midrule

\multicolumn{17}{c}{\textbf{\llamaIIIPLUSBIG}} \\
\addlinespace[2pt]
& \multicolumn{4}{c}{\textbf{(1.4k, 0.2k)}}
& \multicolumn{4}{c}{\textbf{(114k, 5k)}}
& \multicolumn{4}{c}{\textbf{(90k, 8k)}}
& \multicolumn{4}{c}{\textbf{(90k, 8k) Equal Batch}} \\
\cmidrule(lr){2-5}\cmidrule(lr){6-9}\cmidrule(lr){10-13}\cmidrule(lr){14-17}
\textbf{System}
& TTFT (s) & TPS (×A100) & Tok/J & BS
& TTFT (s) & TPS (×A100) & Tok/J & BS
& TTFT (s) & TPS (×A100) & Tok/J & BS
& TTFT (s) & TPS (×A100) & Tok/J & BS \\
\midrule
A100
  & 0.78  & 1.00x  & 1.00x & 256
  & 43.18 & 1.00x  & 1.00x  & 4
  & 29.67 & 1.00x  & 1.00x  & 4
  & 29.67 & 1.00x  & 1.00x  & 4 \\
A100 QuaRot~\cite{QuaRot}
  & 1.17 & 1.08x & 1.08x & 512
  & 42.89 & 1.13x & 1.13x & 8
  & 32.17 & 1.13x & 1.13x & 8
  & 27.69 & 1.11x & 1.11x & 4 \\
H100
  & 0.34 & 2.34x & 1.34x & 256
  & 14.30 & 2.13x & 1.21x & 4
  & 10.10 & 2.04x & 1.22x & 4
  & 10.10 & 2.04x & 1.22x & 4 \\
H100 QuaRot~\cite{QuaRot}
  & 0.44 & 2.36x & 1.35x & 512
  & 16.12 & 2.19x & 1.25x & 8
  & 11.37 & 2.14x & 1.22x & 8
  & 9.88 & 2.08x & 1.18x & 4 \\
TPU v6e
  & 11.7 & 0.85x & N/A & 256
  & 41.96 & 0.46x & N/A & 4
  & 37.61 & 0.47x & N/A & 4
  & 37.61 & 0.47x & N/A & 4 \\
MicroScopiQ~\cite{Microscopiq}
  & 8.32 & 0.79 & 1.59x & 1024
  & 73.28 & 0.20x & 0.41x & 16
  & 49 & 0.17x & 0.35x & 16
  & 23.93 & 0.11x & 0.23x & 4 \\
\textbf{PLENA}
  & 7.58 & 1.82x & \best{3.32x} & 1024
  & 69.10 & \best{2.23x} & \best{4.07x} & 16
  & 43.43 & \best{2.21x} & \best{4.04x} & 16
  & 21.68 & 1.34x & \best{2.45x} & 4\\
\midrule

\multicolumn{17}{c}{\textbf{GPT-OSS 20B (MoE)}} \\
\addlinespace[2pt]
& \multicolumn{4}{c}{\textbf{(1.4k, 0.2k)}}
& \multicolumn{4}{c}{\textbf{(114k, 5k)}}
& \multicolumn{4}{c}{\textbf{(90k, 8k)}}
& \multicolumn{4}{c}{\textbf{(90k, 8k) Equal Batch}} \\
\cmidrule(lr){2-5}\cmidrule(lr){6-9}\cmidrule(lr){10-13}\cmidrule(lr){14-17}
\textbf{System}
& TTFT (s) & TPS (×A100) & Tok/J & BS
& TTFT (s) & TPS (×A100) & Tok/J & BS
& TTFT (s) & TPS (×A100) & Tok/J & BS
& TTFT (s) & TPS (×A100) & Tok/J & BS \\
\midrule
A100
  & 1.46 & 1.00x  & 1.00x  & 1024
  & 11.81 & 1.00x  & 1.00x  & 8
  & 8.05 & 1.00x  &  1.00x & 8
  & 8.05 & 1.00x  & 1.00x  & 8 \\
H100
  & 4.03 & 0.89x  & 0.51x & 1024
  & 1.85 & 3.10x & 1.78x & 8
  & 1.38 & 2.90x & 1.66x & 8
  & 1.38 & 2.90x & 1.66x & 8\\
\textbf{PLENA}
  & 13.41 & \best{1.15x} & \best{2.10x} & 4096
  & 47.63 & 1.96x & \best{3.58x} & 64
  & 41.08 & 1.93x & \best{3.52x} & 64
  & 9.77 & 0.99x & \best{1.79x} & 8 \\
\midrule

\multicolumn{17}{c}{\textbf{Qwen3-32B}} \\
\addlinespace[2pt]
& \multicolumn{4}{c}{\textbf{(1.4k, 0.2k)}}
& \multicolumn{4}{c}{\textbf{(114k, 5k)}}
& \multicolumn{4}{c}{\textbf{(90k, 8k)}}
& \multicolumn{4}{c}{\textbf{(90k, 8k) Equal Batch}} \\
\cmidrule(lr){2-5}\cmidrule(lr){6-9}\cmidrule(lr){10-13}\cmidrule(lr){14-17}
\textbf{System}
& TTFT (s) & TPS (×A100) & Tok/J & BS
& TTFT (s) & TPS (×A100) & Tok/J & BS
& TTFT (s) & TPS (×A100) & Tok/J & BS
& TTFT (s) & TPS (×A100) & Tok/J & BS \\
\midrule
A100
  & 0.88 & 1.00x & 1.00x  & 1024
  & 28.90 & 1.00x  & 1.00x  & 8
  & 19.19 & 1.00x  & 1.00x  & 8
  & 19.19 & 1.00x  & 1.00x  & 8 \\
H100
  & 1.19 & 2.13x & 1.22x & 1024
  & 9.24 & 2.29x & 1.31x & 8
  & 6.29 & 2.21x & 1.26x & 8 
  & 6.29 & 2.21x & 1.26x & 8  \\
\textbf{PLENA}
  & 4.38 & 1.40x & \best{2.56x} & 4096
  & 108.1 & 1.22x & \best{2.23x} & 64
  & 90.71 & 1.23x & \best{2.25x} & 64
  & 23.14 & 1.14x & \best{2.08x} & 8 \\
\bottomrule
  \end{tabular}
\end{table*}

\subsection{System Performance Analysis}

\begin{table}[t]
\centering
\caption{Token usage (prefill/output) across benchmarks:
GSM8K~\cite{gsm8k},
BFCL-Web Search Base~\cite{bfcl},
OSWorld LibreOffice (OSWorld-L)~\cite{osworld}.}
\label{tab:benchmark_token_usage}
\setlength{\tabcolsep}{3pt}
\renewcommand{\arraystretch}{0.9}
\small
\begin{tabular}{lccc}
\toprule
 & GSM8K & BFCL-W & OSWorld-L  \\
\midrule
Prefill (Tokens) & 1.4k & 114k & 90k  \\
Output (Tokens) & 0.2k & 5k   & 8k  \\
\bottomrule
\end{tabular}
\vspace{-10pt}
\end{table}

The system-level performance comparison is shown in \Cref{tab:system_level_combined_reformat}, evaluating both small and large GQA-based LLaMA models as well as the recently published MoE-based GPT-OSS model and Qwen3-32B and supporting long-context inputs. The performance results for PLENA and MicroScopiQ are obtained using our transactional simulator, modeling performance in a 7 nm technology node. For fairness, we conduct a system-level comparison against a 4$\times$A100 SXM GPU system (80\,GB HBM and 1.99\,TB/s bandwidth per GPU), a 4$\times$H100 SXM GPU system (80\,GB HBM and 3.35\,TB/s bandwidth per GPU), and a 16$\times$TPU v6e system (32\,GB HBM and 1.56\,TB/s bandwidth per device). Both PLENA and MicroScopiQ are modeled as 16-accelerator systems with aggregate HBM capacity and bandwidth equivalent to the TPU system. To account for GPUs' non-compute components, the number of devices is determined by approximately aligning multiplier counts rather than silicon area. The co-design-selected PLENA configuration—(\texttt{BLEN} = 32, \texttt{MLEN} = 2048, \texttt{VLEN} = 2048, Precision W/A/KV = 4/4/4)—demonstrates improved performance across all evaluated workloads.

As shown, PLENA achieves higher TPS than both the A100 and TPU v6e under identical HBM settings and multiplier counts, reaching up to 2.23$\times$that of the A100 and 4.70$\times$ that of the TPU v6e for agentic workload. The higher TTFT observed in PLENA is explained by its ability to store more batches within the same HBM capacity using our quantization scheme. As batch size increases, the prefill stage grows longer due to additional memory accesses and computation.

\section{Conclusion}

We present \textbf{PLENA}, a novel accelerator design system that exploits a flattened systolic array for efficient agentic inference acceleration.
We identify the underutilization challenges posed by memory bandwidth and capacity walls for agentic model inference.
In order to address them, we propose an asymmetric quantization scheme for hardware acceleration in MX and native architectural support for FlashAttention.
Beyond the hardware, PLENA also presents a full system exploration framework, including new ISA support, automated code generation, multi-level simulators, and a co-design exploration engine.
This provides an exploration platform beyond a specific accelerator architecture implementation and enables future research to prototype and explore optimizations for emerging transformer models, similar to Berkley's GEMMNI framework for DNNs~\cite{gemmini-dac}.
Our future work will focus on integrating PLENA with GPU systems for heterogeneous LLM acceleration, combining the best of both worlds.


\bibliographystyle{IEEEtranS}
\bibliography{refs}

\end{document}